\documentclass[aps,amsmath,amssymb,amsthm,dsfont,prx,nofootinbib,twocolumn,superscriptaddress,nobibnotes]{revtex4-2}
\usepackage{times}
\usepackage[colorlinks=true,citecolor=blue,linkcolor=blue]{hyperref}
\usepackage{graphicx,epstopdf}              
\usepackage{mathtools}                        
\usepackage{amsfonts,amsthm,amssymb,latexsym,amsmath}
\usepackage{mathrsfs}
\usepackage{bbm}
\usepackage{units}  
\usepackage{textcomp} 
\usepackage{hhline}
\usepackage{multirow}   
\usepackage{makecell}
\usepackage{tabularx}
\usepackage{booktabs}
\usepackage{hyperref}
\hypersetup{
colorlinks = true,
linkcolor=magenta,
citecolor=[rgb]{0.15,0.65,0.13},
urlcolor = [rgb]{0.25, 0.41, 0.88}}
\usepackage{float}
\usepackage{cleveref}
\usepackage{tikz}
\usepackage{booktabs}
\usepackage{cancel}
\usepackage{colortbl}

\def\bra#1{\mathinner{\langle{#1}|}}
\def\ket#1{\mathinner{|{#1}\rangle}}

\def\inner#1{\mathinner{\langle{#1}\rangle}}
\def\bs#1{\boldsymbol{#1}}

\def\ZZ{\mathbb Z}

\def\cX{\mathcal X}
\def\cZ{\mathcal Z}

\newcommand{\nablapic}{\raisebox{.0\height}{\includegraphics[scale=0.1]{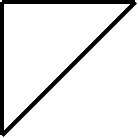}}}

\newcommand{\TQFT}{$\mathrm{TQFT}_{4}$}

\makeatletter
\def\l@subsection#1#2{}
\def\l@subsubsection#1#2{}
\makeatother
\begin{document}

\title{Hybrid Fracton Phases: Parent Orders for Liquid and Non-Liquid Quantum Phases}

\author{Nathanan Tantivasadakarn}
\email{ntantivasadakarn@g.harvard.edu}
\affiliation{Department of Physics, Harvard University, Cambridge, Massachusetts 02138, USA}
\author{Wenjie Ji}
\email{wenjieji@ucsb.edu}
\affiliation{Department of Physics, University of California, Santa Barbara, California 93106, USA}
\affiliation{Department of Physics, Massachusetts Institute of Technology, Cambridge, Massachusetts 02139, USA}
\author{Sagar Vijay}\email{sagar@physics.ucsb.edu}
\affiliation{Department of Physics, University of California, Santa Barbara, California 93106, USA}

\begin{abstract}

We introduce hybrid fracton orders: three-dimensional gapped quantum phases that exhibit the phenomenology of both conventional three-dimensional topological orders and fracton orders.  Hybrid fracton orders host both (i) mobile topological quasiparticles and loop excitations, as well as (ii) point-like topological excitations with restricted mobility, with non-trivial fusion rules and mutual braiding statistics between the two sets of excitations. Furthermore, hybrid fracton phases can realize either conventional three-dimensional topological orders or fracton orders after undergoing a phase transition driven by the condensation of certain gapped excitations. Therefore they serve as parent orders for both long-range-entangled quantum liquid and non-liquid phases. We study the detailed properties of hybrid fracton phases through exactly solvable models in which the resulting orders hybridize a three-dimensional $\ZZ_2$ topological order with (i) the X-Cube fracton order, or (ii) Haah's code.  The hybrid orders presented here can also be understood as the deconfined phase of a gauge theory whose gauge group is given by an Abelian global symmetry $G$ and subsystem symmetries of a normal subgroup $N$ along lower-dimensional sub-regions. A further generalization of this construction to non-Abelian gauge groups is presented in \cite{TJV2}.
\end{abstract}
\maketitle
  \hypersetup{linkcolor=blue}
\tableofcontents
\hypersetup{linkcolor=magenta}
\section{Introduction}

Gapped fracton phases of matter  \cite{Chamon2005,Haah2011,Yoshida2013,VijayHaahFu2015,VijayHaahFu2016} are quantum phases characterized by the presence of fractionalized, point-like excitations with highly restricted mobility, and a robust ground-state degeneracy that can grow sub-extensively with system size, due to an intricate pattern of long-range entanglement. This phenomenology is in stark contrast to that of more familiar, topologically ordered phases that can host mobile point-like excitations with nontrivial self-statistics and mutual-statistics in two spatial dimensions, along with loop excitations in three dimensions.  Gapped fracton orders appear in two broad categories: Type I orders, such as the X-Cube  fracton order \cite{VijayHaahFu2016}, host both immobile quasiparticles (fractons) as well as fractionalized excitations with restricted mobility, while in Type II orders, such as Haah's code \cite{Haah2011}, all fractionalized excitations are immobile, and cannot be separated without incurring an energy cost.  More exotic non-Abelian fracton orders have been recently explored, in which certain excitations with restricted mobility have a protected internal degeneracy \cite{Vijay2017generalization, SongPremHuangMartinDelgado19, PremHuangSongHermele2019,WilliamsonCheng20}, analogous to the quantum dimension of non-Abelian quasiparticles in two-dimensional topological phases. More possibilities of immobile excitations, such as immobile string excitations have also began to be explored \cite{li2020fracton}.

Progress has been made towards understanding some universal properties of fracton orders that are characteristic of the phase, such as fusion and braiding processes for fractionalized excitations \cite{SlagleKim2017,PaiHermele2019}, topological entanglement entropy in the ground-state  \cite{MaSchmitzParameswaranHermeleNandkishore2018,ShiLu2018,HeZhengBernevigRegnault2018}, and the foliated structure of certain Type I fracton orders which allows these   phases to easily ``absorb" two-dimensional topological orders through the action of a finite-depth quantum circuit \cite{ShirleySlagleWangChen2018,ShirleySlagleChen19,ShirleySlagleChen19_2}, in contrast to a conventional \emph{quantum liquid}, which can similarly absorb short-range-entangled degrees of freedom \cite{ZengWen15}.  Quantum field theories that capture the  low-energy properties of Type I fracton orders have been recently proposed \cite{SlagleKim2017, Slagle20,  SeibergShao2020_3, SlagleAasenWilliamson19,  GorantlaLamSeibergShao20_1, GorantlaLamSeibergShao20_2}.  

Though fracton orders have attracted intense study, they have so far been treated as exotic non-liquid phases that stand alone from  conventional, three-dimensional topological orders, which can be described at low energies by topological quantum field theories in $(3+1)$ spacetime dimensions (\TQFT). Some indirect relations between fracton orders and conventional quantum liquids have been identified.  First, strongly coupled stacks of lower-dimensional topological phases can realize certain fracton orders or three-dimensional topological orders, independently  \cite{WangSenthil13,JianQi14,Vijay2017,MaLakeChenHermele2017,SlagleKim17_2,PremHuangSongHermele2019,Fuji19,ShirleySlagleChen20,Schmitz19,WilliamsonDevakul20_1,WilliamsonDevakul20_2,Wen20,Wang20}. Second, lattice models in which the gapped excitations contain non-Abelian fractons as well as mobile particles which behave similarly to the charges in a three-dimensional $D_{4}$ gauge theory, have been recently proposed  \cite{BulmashBarkeshli2019,PremWilliamson2019,StephenGarre-RubioDuaWilliamson2020,AasenBulmashPremSlagleWilliamson20}.  However, key properties of these models including (i) their relationship to conventional quantum liquid orders such as the $D_{4}$ gauge theory and (ii) the braiding and fusion of excitations, are not fully understood. Whether properties of certain long-range-entangled quantum liquid and non-liquid states can coexist, or be possibly unified into a ``parent" order, has remained an open question. 

In this work, we answer this question directly, by proposing a family of \emph{hybrid} fracton orders, which host both the exotic excitations of a fracton phase, as well as the point- and loop-like excitations that appear in a \TQFT.  Within these hybrid phases, the two kinds of excitations have non-trivial mutual statistics and fusion rules -- i.e. collections of excitations native to the fracton order can fuse into excitations of the quantum liquid order and vice versa -- so that the hybrid phase is truly distinct from a tensor product of the two orders.  These hybrid orders further serve as clear parent phases for both conventional  three-dimensional topological orders and Type I or Type II fracton orders, since they can realize either order after undergoing a phase transition in which an appropriate set of gapped  excitations condense.   After introducing a framework for understanding the emergence of these orders, we concretely characterize certain hybrid ordered phases through a series of exactly solvable models, which provide a theoretical toolbox to determine the topological data -- including fusion, braiding, and the mobilities of excitations -- in full detail.

A number of outstanding questions about these phases that we introduce, and their generalizations, remain to be addressed.  First, it remains to be understood whether hybrid fracton phases can fit into the existing framework of foliated fracton orders. For example, can a two-dimensional topological order be ``exfoliated" from Type I hybrid fracton models? This would also clarify their entanglement renormalization group flow, which exhibits dramatically different behaviors between liquid and non-liquid phases  \cite{Haah14,DuaSarkarWillamsonCheng20}. Whether hybrid fracton phases can quantitatively improve upon the performance of existing quantum memories based on Type II fracton orders \cite{BravyiHaah2013} also remains to be studied.  Furthermore, a field-theoretic understanding of such orders could shed light on the universal properties of these states at low energies, other proximate phases, and other hybridizations of liquid and non-liquid orders that are possible. One such construction has been recently realized in Ref. \onlinecite{HsinSlagle21}. We note that a systematic study of more general hybrid fracton orders, which also yield non-Abelian fracton excitations, have been presented in recent follow-up work \cite{TJV2}. %We hope to address this last question in future work \cite{TJV2}. 

\begin{figure}[t!]
\centering
$\begin{array}{ccc}
    \raisebox{-.5\height}{\includegraphics[scale=0.4]{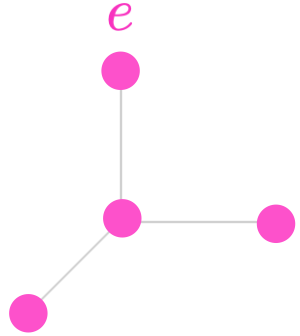}}
    & & \raisebox{-.5\height}{\includegraphics[scale=0.45]{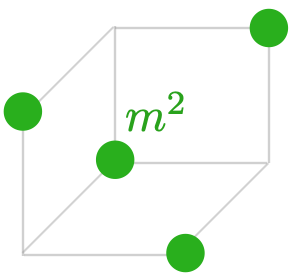}}\\
    \text{(a)} & & \text{(b)}\\
    \\
    \raisebox{-.5\height}{\includegraphics[scale=0.43]{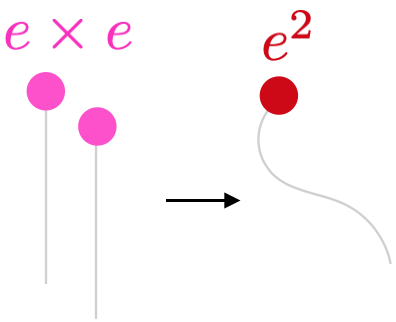}} & & \raisebox{-.5\height}{\includegraphics[scale=0.31]{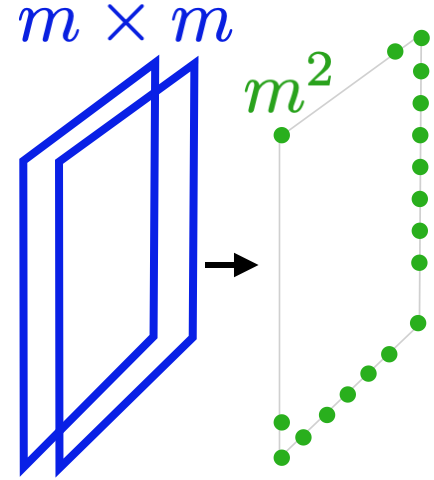}}\\
    \text{(c)} & & \text{(d)}\\
    \\
    \raisebox{-.5\height}{\includegraphics[scale=0.59]{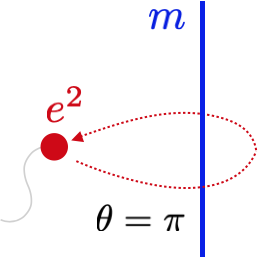}}
    & &\raisebox{-.5\height}{\includegraphics[scale=0.57]{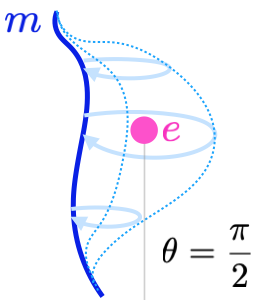}}\\
    \text{(e)} & & \text{(f)}\\
\end{array}$
    \caption{{\bf Hybrid Haah's Code:} In the hybrid Haah's code, the two species of fracton excitations ($e$ and $m^{2}$) that are native to Haah's code \cite{Haah2011} are created in the geometric arrangements shown in (a) and (b).  Pairs of $e$ fractons fuse into a mobile $\ZZ_{2}$ charge ($e^{2}$) as in (c).  The hybrid order also hosts a flux loop $m$. Two identical, rectangular flux loops fuse to generate an arrangement of the $m^{2}$ fracton excitations, shown schematically in (d).  The precise geometry of the generated fracton excitations is presented in Fig. \ref{fig:mfusionHaah} in Sec. \ref{Z4Z2Haah}. The $e^{2}$ charge has identical mutual statistics with a flux loop $m$ as in the 3d toric code as shown in (e). In (f), braiding the flux loop $m$ around the $e$ fracton gives a non-trivial  phase that is consistent with the fusion and braiding in (c) and (e).}
    \label{fig:Haah_Hybrid}
\end{figure}

\begin{table}
\caption{{\bf Fusion in Hybrid Fracton Orders:} In the hybrid orders studied in this work, the excitations $e^{2}$ $(m)$ resemble the charge (flux) in the 3d $\ZZ_2$ toric code, while $e$ and $m^{2}$ resemble the  two species of excitations with reduced mobility in a $\ZZ_2$ fracton order, respectively. These excitations and their composites form all of the gapped excitations in the hybrid order, and some of their characteristic fusion rules are shown below. The inverses of  $e$ and $m$ -- labeled $\bar{e}$ and $\bar{m}$, and defined by the relation $\bar{e}\times e = \bar{m}\times m = 1$ -- have identical mobilities as $e$ and $m$, respectively.  \\}  %Excitations with restricted mobility can fuse into mobile charges, while flux loops can fuse into a number of point excitations $m^{2}$ with restricted mobility of the fracton order, as explained in the text.}
\def\arraystretch{1.3}
\begin{tabular}{| c c |} 
\hline 
&{\bf Excitations \& Fusion Rules}\\
\hline
&Generating Set $=\{1,e,e^{2},m,m^{2}\}$\\
&$e^{2} \equiv e\times e =$ mobile $\ZZ_{2}$ charge\\
&$m =$ flux loop\\
&$m^{2}\times m^{2}  = e^{2}\times e^{2} = 1$\\
\hhline{==}
&{\bf Hybrid Toric Code Layers (Sec. \ref{1foliatedZ4Z2})} \\
&$e \equiv$ planon \hspace{.2in} $m^2 \equiv$ planon\\
&$m\times m = $ planons ($m^{2}$) along loop\\
\hline
&{\bf Fractonic Hybrid X-Cube (Sec. \ref{3foliatedZ4Z2})} \\
&$e \equiv$ fracton \hspace{.2in} $m^2 \equiv$ lineon\\
&$m\times m ~=$ lineons ($m^{2}$) at corners \\
\hline
&{\bf Lineonic Hybrid X-Cube (Sec. \ref{sec:Z42Z22lineon})}  \\
&$e \equiv$ lineon \hspace{.2in} $m^2 \equiv$ fracton\\
&$m\times m ~=$  fractons ($m^{2}$) at corners \\
\hline
&{\bf Hybrid Haah's Code (Sec. \ref{Z4Z2Haah})}  \\
&$e \equiv$ fracton \hspace{.2in} $m^2 \equiv$ fracton\\
&$m\times m ~=$ fractons ($m^{2}$) along loop\footnote{The precise geometric arrangement of the fractons generated by the loop fusion is presented in Sec. \ref{Z4Z2Haah}.} \\
\hline
\end{tabular}
\label{tab:summary}
\end{table}

{\bf \emph{Summary of Main Results: }}We now provide a  detailed summary of our main results, and an outline of this work. 

 To illustrate the properties of hybrid fracton orders, we introduce four exactly solvable models of these hybrid phases, in increasing levels of complexity. All of the models introduced can be thought of as a hybrid of a (liquid) $\ZZ_2$ toric code in three dimensions, and a (non-liquid) $\ZZ_2$ fracton model, due to the fact that the hybrid order hosts both the gapped excitations of the toric code, as well as the exotic excitations of the fracton order.   In fact, the hybrid orders we introduce have the same ground state degeneracy as the tensor product of the two orders on the three-torus.  Nevertheless, the hybrid phases differ from a trivial tensor product due to the non-trivial fusion and braiding of the gapped excitations.   
 
 The excitations in the hybrid orders that we consider, along with some of their braiding and fusion rules, may be summarized succinctly.  All of these orders host a mobile $\ZZ_2$ charge (labeled $e^{2}$) and a $\ZZ_2$ flux loop (labeled $m$), which have the same mobility and mutual statistics as the $\ZZ_2$ charge and flux in the 3d toric code. Furthermore, the hybrid order hosts an excitation with restricted mobility (labeled $e$) and its conjugate excitation (labeled $m^{2}$) which are in correspondence with the excitations in a particular Type I or Type II fracton order.  For example, $e$ can correspond to the fracton excitation in the X-Cube model, with $m^{2}$ then corresponding to the conjugate excitation in the X-Cube phase which is only mobile along lines (the lineon).  The resulting hybrid phase is then termed the fractonic hybrid X-Cube order, where ``fractonic" refers to the mobility of the $e$ excitation.  In select situations, this labeling is unnecessary as the hybrid order can be defined unambiguously.  In Haah's code, for example, both species of excitations are fractons, which are further exchanged by a duality transformation \cite{Haah14}. Therefore, in constructing a ``hybridization" of Haah's code with the 3d toric code, choosing $e$ to be either of the fracton excitations in Haah's code yields the same hybrid order, whose properties are summarized schematically in Fig. \ref{fig:Haah_Hybrid}. 
 
 Our labeling of the gapped excitations in the hybrid fracton orders is suggestive of their fusion rules, which are summarized in Table \ref{tab:summary}. These fusion rules yield new phenomena that are not separately possible in either the fracton or three-dimensional toric code orders. For example, in both the hybrid Haah's code and in the fractonic hybrid X-Cube order, two fractons $(e\times e)$ fuse into a completely \emph{mobile} quasiparticle $(e^2)$. This is particularly striking, as a single fracton is completely immobile, and collections of these fractons can only form excitations with significantly reduced mobility in a Type I fracton order. Additionally, the fusion of a pair of loop excitations $(m\times m)$ yields a geometric pattern of $m^{2}$ excitations, which are fractons in the hybrid Haah's code.   
 
 Apart from the fusion rules, we obtain a universal braiding phase for two excitations  $e^{a}$ and $m^{b}$,  when at least one of the two excitations exhibits enough mobility to remotely detect the other.  This braiding process leads to the accumulation of a universal phase $e^{i\theta_{ab}}$ where 
 \begin{align}
     \theta_{ab} = \frac{i\pi}{2}ab,
 \end{align}
 and with $a$, $b\in\{0,1,2,3\}$, in all of the hybrid orders that we present.  Other braiding processes that are specific to each hybrid order are also studied, which are not summarized here.  

The emergence of hybrid fracton orders may be more generally understood in two complementary ways.  First, our hybrid orders can be obtained by starting with a Type I or Type II fracton order which is enriched by an on-site Abelian global symmetry (e.g.  $\ZZ_2$), so that certain excitations of the fracton order carry fractional quantum numbers under the symmetry.  Gauging this global symmetry then yields a hybrid fracton order, in which certain excitations of the original fracton order can fuse into the gauge charge in a conventional topological order (e.g. the gapped charge in a $\ZZ_2$ gauge theory).  

Equivalently, the hybrid order can be thought of as the deconfined phase of a gauge theory.  We may start with a short-range-entangled (SRE) quantum system with global symmetry $G$ and subsystem symmetries $N$, where $N$ is a normal subgroup of $G$; the subsystem symmetries are defined as symmetry transformations along extensive sub-regions of the lattice (e.g. planes).  Importantly, the subsystem and global symmetries are not independent of each other, and their interplay is such that the gapped,  symmetric excitations of the SRE phase  can  be  (i)  charged  under  the  global symmetry or (ii) charged under a combination of planar symmetries and the global symmetry, so that gauging these symmetries yields a hybrid fracton order. As a consequence of this construction, we also refer to the  hybrid order as a $(G,N)$ gauge theory, and in this work we restrict our attention to Abelian groups $G$ and $N$, where $G/N = \ZZ_2$. Because the hybrid order is the deconfined phase of a gauge theory, we will often refer to its gapped excitations as charges or fluxes depending on whether the excitation is related to a (i) gapped, symmetric excitation in the ungauged, SRE phase which transforms under the symmetry group (charge) or (ii) a defect of the symmetry group (flux).  $(G,N)$ gauge theories for more general groups $G$ and $N$ are studied extensively in a follow-up work \cite{TJV2}, where it is found that gauging the Abelian global symmetry $G$ and subsystem symmetries $N$ of a SRE state yields a hybrid order that hybridizes a 3d $G/N$ toric code and a  fracton model based on the subsystem symmetry gauge group $N$. 

We now provide an outline of this work.  In Sec. \ref{1foliatedZ4Z2}, we introduce the simplest example of a hybrid phase, which hybridizes the order in a stack of two-dimensional (2d) $\ZZ_2$ toric codes and the three-dimensional (3d) $\ZZ_2$ toric code.  This order -- termed the hybrid toric code layers -- can be obtained either as a  generalized gauge theory, or by condensing a set of gapped excitations in a stack of 2d $\ZZ_4$ toric codes.  These two complementary ways of obtaining the hybrid order provide an important understanding about the fusion and braiding statistics of the gapped excitations.  The intuition obtained from this example extends to the hybrid fracton orders that we consider subsequently.   

In the remaining sections, we introduce more complex hybrid phases that hybridize a fracton order with a 3d toric code topological order.  For the  hybrid Type I fracton models that we present, we choose the X-Cube order \cite{VijayHaahFu2016} as our input. In this case, there are two possible hybrid orders that can be obtained, if the gapped excitation $e$ is chosen to be the fracton in the X-Cube model or the lineon excitation. The former yields the fractonic hybrid X-Cube order, in which a pair of fractons fuse to the mobile charge $e^{2}$, and is introduced in Sec. \ref{3foliatedZ4Z2}.  The latter case, where a pair of lineons fuse to the mobile charge, is presented in Sec. \ref{sec:Z42Z22lineon}. The equivalance between the ground-state degeneracy of the fractonic hybrid X-Cube order and of the tensor product of the X-Cube and toric code orders on the three-torus is related to an isomorphism between the algebra of closed Wilson loop and membrane operators in the ground-state subspace of these orders, which we identify. Hybrid Type II orders can also exist\footnote{Unlike a type II fracton order, in which all topological excitations are strictly immobile, a \emph{hybrid} Type II order hosts the excitations of both a liquid order and a type II fracton order, and can have mobile topological excitations.} and we introduce a hybrid of the 3d $\ZZ_2$ toric code and Haah's code \cite{Haah2011} in Sec. \ref{Z4Z2Haah} and study its properties in detail. 

Lastly, in Sec. \ref{sec:parent}, we study the proximate phases of the hybrid fracton orders, which establishes that these models are parent states for both conventional topological orders, as well as fracton orders.   We explicitly demonstrate that for either the Type I or Type II hybrid fracton orders that we introduce, condensing an appropriate set of gapped excitations can drive a phase transition into either a $\ZZ_2$ topologically ordered phase or a $\ZZ_2$ fracton phase.  We show that the phase transition between one of the hybrid orders and an X-Cube fracton order can be direct and \emph{continuous}, and related to the Higgs transition in a three-dimensional $\ZZ_2$ gauge theory in a particular limit, though the generic nature of this phase transition remains to be understood.

Interestingly, we find that a common feature of the hybridized model is that they can be thought of as promoting certain $\ZZ_2$ degrees of freedom in the tensor product of a liquid and non-liquid order into $\ZZ_4$ degrees of freedom. More concretely, starting from a product of the $\ZZ_2$ toric code and a $\ZZ_2$ fracton model, the hybridization can be viewed as pairing up qubits of the toric code with qubits of the fracton model, and promoting these pairs to a $\ZZ_4$ qudit. For cases where the degrees of freedom of the two models both live on edges, such as the models in Secs. \ref{1foliatedZ4Z2} and \ref{sec:Z42Z22lineon}, we are able to rewrite the Hamiltonian as a mix of $\ZZ_2$ qubits and $\ZZ_4$ qudits. For those in Secs. \ref{3foliatedZ4Z2} and \ref{Z4Z2Haah}, the positions of the degrees of freedom of the toric code and fracton model do not match, and the algebra of operators in the hybrid model is more involved (see Eqs. \eqref{equ:xiezetapalg} and \eqref{equ:xiezetapalgHaah}).

 In Appendix \ref{app:2subsystemdef}, we give a self-contained discussion of the definition of an Abelian $(G,N)$ symmetry, and a qualitative description of the process of gauging such a symmetry. A more general construction of hybrid fracton models involving a general finite group $G$ is presented in a follow-up work \cite{TJV2}.

\section{Hybrid Toric Code Layers}\label{1foliatedZ4Z2}
\begin{figure}[t!]
    \centering
    \includegraphics[scale=0.25]{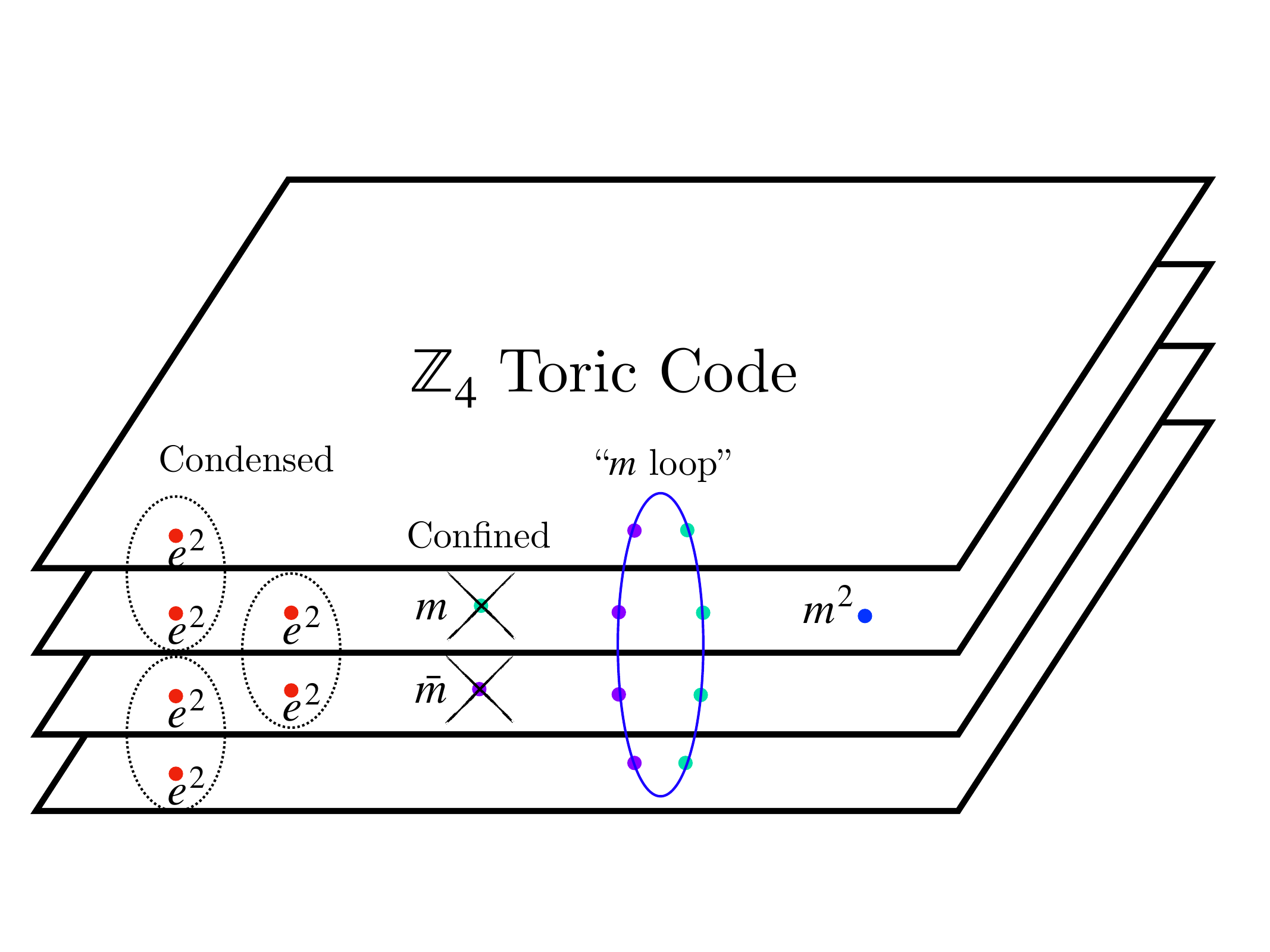}
    \caption{ {\bf Hybrid Toric Code Layers from a stack of $\ZZ_4$ Toric Codes:} An alternate construction of the hybrid toric code layers, which clarifies the nature of the flux excitations is shown schematically. Starting with a stack of $\ZZ_4$ toric codes, we condense pairs of $e^2$ anyons in adjacent layers. The anyons $m$ and $\bar m$ in each layer is confined, but a composite excitation composed of $m$ and $\bar m$ pairs in each layer -- the ``$m$ loop"-- braids trivially with the condensate and therefore remains as a topological excitation.}
    \label{fig:Z4TCcondensation}
\end{figure}

In this section, we begin by describing the simplest construction of a hybrid order, which hybridizes a stack of 2d toric code layers with a 3d toric code, as  described by the Hamiltonian (\ref{equ:1foliatedHam}).  The resulting hybrid order contains the excitations of both the 2d toric code, which are restricted to move within planes, as well as those of the 3d toric code, with non-trivial braiding and fusion rules.

We first obtain this hybrid order by gauging the symmetries of a short-range-entangled (SRE) phase.  We start with $L$ independent copies of a two-dimensional, SRE states, each with a global $\ZZ_4$ symmetry, and in a trivial gapped, symmetry-preserving (paramagnetic) state. The full symmetry group of the stacked layers is $\ZZ_4^L$. The excitations in each layer that are charged under this symmetry group (charges) can be labeled by an integer modulo 4, and cannot move across layers.

\begin{table*}[t!]
    \centering
        \caption{{\bf Excitations in the hybrid toric code layers:} A summary of the pure charge and flux excitations in the hybrid toric code layers is provided below, along with the local operators that measure these excitations in the lattice model Eq. \eqref{equ:1foliatedHam}.}
    \def\arraystretch{1.5}
    \begin{tabular}{|c|l|l|l| }
    \hline
    Excitation & \multicolumn{1}{c|}{Creation operator} & \multicolumn{1}{c|}{Charge}  & \multicolumn{1}{c|}{Local Wilson operator} \\
    \hline
 planon $e$ & End points of $\cZ$ on $x,y$ edges & $\bs A_v = i, \bs A_{v}^{2d}=-1$ & $\bs B_{p_{xy}}$ = closed $e$ loop around $p_{xy}$ \\
 \hline
\multirow{2}{*}{mobile charge $e^2$} & End points of $\cZ$ on  $x,y$ edges  & \multirow{2}{*}{$\bs A_v = -1$} & \multirow{2}{*}{$\bs B_{p}^{3d}$ = closed  $e^2$ loop around $p$} \\
&End points of $Z$ on  $z$ edges &&\\
\hline
\multirow{2}{*}{loop $m$} & Boundary of $X$ membrane in $xy$, plane  & $\bs B_{p}^{3d}= -1$ & \multirow{2}{*}{$\bs A_v$ = closed $m$ membrane around $v$}  \\
 &  Boundary of $\cX$ membrane in $xz$, $yz$ plane &   $\bs B_{p}^{3d}= -1$, $\bs B_{p_{xy}}= \pm i$ &  \\
 \hline
planon $m^2$ & End points of $\cX^2$ in $xy$ plane  &$\bs B_{p_{xy}}= -1$ &$\bs A_v^{2d}$ = closed $m^2$ loop around $v$\\
        \hline
    \end{tabular}
    \label{tab:Z41foliated}
\end{table*}

Next, we may break the $\ZZ_4^L$ symmetry by adding a coupling between adjacent layers that allows charge-2 excitations to tunnel between layers. We observe that since the charge in each layer is only now conserved modulo 2, the $\ZZ_4$ symmetry in each layer is now broken down to $\ZZ_2$. Nevertheless, the global $\ZZ_4$ symmetry defined as the diagonal $\ZZ_4$ symmetries of all layers is still preserved. 

The charge excitations in the SRE state still transform as a $\ZZ_4$ representation under the global symmetry, so they can still be labeled by an integer modulo 4. However, they exhibit mobility constraints due to the fact that they transform under the remaining $\ZZ_2$ planar symmetry in each layer. The even charges transform trivially under the $\ZZ_2$ planar symmetry, meaning they are fully mobile, while the odd charges are also charged under the planar symmetry, meaning they can only move within their respective planes. From this, it is also apparent that a fusion of two identical odd charges results in a fully mobile even charge.

We now gauge all of the symmetries of this model. The properties of the charge excitations of the SRE state carry over to the gauge charges of the resulting hybrid order. Qualitatively, we can first gauge the $\ZZ_2$ planar symmetries, which creates stacks of $\ZZ_2$ toric codes. The global symmetry is now reduced from $\ZZ_4$ to $\ZZ_2$ because we have also gauged a $\ZZ_2$ subgroup of the global symmetry, which is the product of the $\ZZ_2$ planar symmetry in every layer. As a consequence, the stack of toric codes are each enriched by the remaining global $\ZZ_2$ symmetry. In particular, the global symmetry fractionalizes on the toric code anyon $e$ in every layer. Finally, we may also gauge the global $\ZZ_2$ symmetry to obtain the desired hybrid model.

While the above construction is well-defined, it sheds less light on the nature of the flux excitations in the final hybrid model. To study the flux excitations, we find it more insightful to consider the following alternate route, which will result in the same hybrid order. We start with the SRE state and temporarily neglect the interlayer couplings, so that each layer has the full $\ZZ_4$ symmetry.  By gauging the $\ZZ_4$ symmetry in each layer, we obtain a stack of $\ZZ_4$ toric codes. Then, to restore the interlayer couplings, we condense pairs of $e^2$ anyons of the $\ZZ_4$ toric code between every adjacent layer, as shown in Figure \ref{fig:Z4TCcondensation}.  As a result, the $e^2$ anyons in each layer are all now in the same superselection sector in the condensate phase, making the $e^2$ particle mobile in the $z$ direction. In addition, the unit flux $m$, which braids non-trivially with the $e^{2}$ pairs is confined, but a composite loop excitation composed of $m$-$\bar m$ pairs in each layer remains deconfined. We will refer to this loop excitation as the ``$m$-loop'', a gauge flux of the hybrid model. The anyon $m^2$, however braids trivially and survives as a well-defined excitation in the condensed phase. It therefore remains as a point particle confined to each layer. 

The considered condensation has interesting consequences in terms of the mobility of the particles under fusion. Identically to the charge excitations before gauging, the gauge charge $e$ is a planon, but fusion with another gauge charge gives $e^2$, a fully mobile excitation. In addition, obtaining the hybrid phase by  condensing excitations in a stack of $\ZZ_4$ toric codes allow us to determine the fusion of the flux excitations. The $m$-loop is fully mobile, but upon fusion with itself, it decomposes into pairs of $m^2$ planons in each layer. The types of excitations and their  mobilities in the hybrid toric code layers are summarized in Table \ref{tab:Z41foliated}.  

We note that other than the unusual mobilities, the \emph{statistics} of the excitations are the same as those of a $\ZZ_4$ toric code model. That is, the mutual statistics of the excitations $e^a$ and $m^b$ for $a=0,1,2,3$ is just $i^{ab}$. While it could be suggestive to think that the final hybrid model simply decouples into a 3d toric code (with mobile excitations $e^2$ and $m$) tensored with a stack of 2d toric codes (formed by planar excitations $e$ and $m^2$) because each of the pairs above has a mutual $-1$ braiding statistics, this can be refuted by noticing the mutual statistics of $i$ between the $m$ loop and the $e$ planon, which cannot occur in the stacked model.

A simple exactly-solvable lattice model for the hybrid toric code layers can be explicitly constructed. The model is a hybrid of the 2d and 3d toric codes. On a square lattice, we place a $\ZZ_2$ qubit on the $z$ links with the usual $Z$ and $X$ Pauli operators, and a $\ZZ_4$ qudit on the $x$ and $y$ links with the following clock and shift operators
\begin{align}
   \cZ &= \sum_{n=0}^{3}i^{n}\ket{n}\bra{n}, \hspace{.2in} \cX = \sum_{n=0}^{3}\ket{n+1}\bra{n},
   \label{equ:clockshift}
\end{align}
which satisfy $\cZ\cX = i\cX\cZ$. The Hamiltonian is given by
\begin{align}
    H_\text{Hybrid} &=H_{TC_3}' + H_{TC_2}', \nonumber\\
    H_{TC_3}' &=-\sum_v \frac{\bs A_v+\bs A_v^\dagger}{2} - \sum_{p} \frac{1 + \bs B_{p}^{3d}}{2},\nonumber \\
    H_{TC_2}' &= -\sum_v \frac{1+\bs A_v^{2d}}{2} 
-  \sum_{p_xy} \frac{\bs B_{p_{xy}}+  \bs B_{p_{xy}}^\dagger}{2}.
    \label{equ:1foliatedHam}
\end{align}
where $p_{xy}$ refer to plaquettes that are in the $xy$ plane only. The explicit form of the operators are
\begin{align}
    \bs A_v &= \raisebox{-0.5\height}{\includegraphics[scale=1]{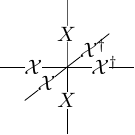}},
\end{align}
\begin{align}
    \bs B_{p}^{3d} &= \raisebox{-0.5\height}{\includegraphics[scale=1]{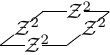}},
    \raisebox{-0.5\height}{\includegraphics[scale=1]{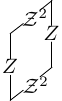}}, \raisebox{-0.5\height}{\includegraphics[scale=1]{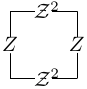}}
\end{align}
\begin{align}
    \bs A_v^{2d} &= \raisebox{-0.5\height}{\includegraphics[scale=1]{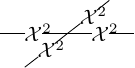}},
\end{align}
\begin{align}
    \bs B_{p_{xy}} &= \raisebox{-0.5\height}{\includegraphics[scale=1]{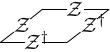}}.
\end{align}
The hybridization of the 2d and 3d $\ZZ_2$ toric codes can be seen from the fact that the edges in the $xy$ plane have been modified from $\ZZ_2$ to $\ZZ_4$ qudits. The two models are coupled in such a way that the vertex term of the 3d toric code $\bs A_v$ squares to the vertex term of the 2d toric code $\bs A_v^{2} = \bs A_v^{2d}$. Furthermore, for plaquettes in the $xy$ plane, the plaquette term of the 2d toric code $\bs B_{p_{xy}}$ squares to the $xy$ plaquette of the 3d toric code $\bs B_{p_{xy}}^{3d}$.

\begin{figure}[t!]
    \centering
    \includegraphics[scale=.73]{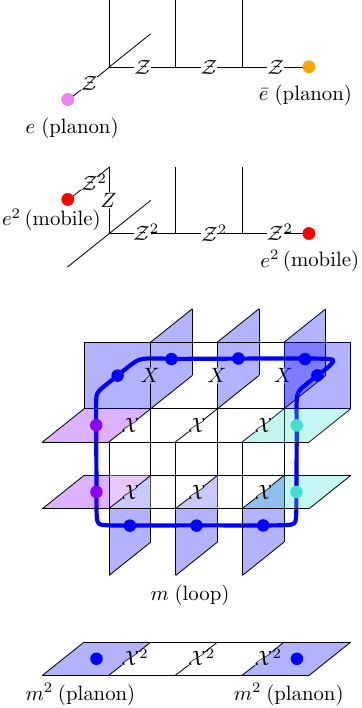}
    \caption{{\bf Geometry of the Excitations in the Hybrid Toric Code Layers:} Excitations of the hybrid toric code layers from the Hamiltonian \eqref{equ:1foliatedHam} are shown. For charge excitations, the colors magenta, red, and orange correspond to $\bs A_v=i,-1,-i$, respectively. For flux excitations, blue refers to $\bs B_{p}^{3d},\bs B_{p,xy}=-1$, while cyan and purple refers to $\bs B_{p_{xy}}=i,-i$, respectively. }
    \label{fig:exc_1foliated}
\end{figure}

The phenomenology of the hybrid model can be readily checked from this lattice model as illustrated in Figure \ref{fig:exc_1foliated} and summarized in Table \ref{tab:Z41foliated}. The planon $e$ corresponds to $\bs A_v = i$, and can only be excited at the end points of a string of $\mathcal Z$ in the $xy$ plane. Squaring this string operator creates the excitation $e^2$ at its end points, which satisfy $\bs A_v =-1$. However, $e^2$ is a mobile particle because it can also hop in the $z$ direction by acting with $Z$ on $z$ edges.

To create the flux loop $m$, we apply $\mathcal X$ on every $x$ or $y$ link, and $X$ on every $z$ link that intersects a surface $\mathcal S'$ on the dual lattice. The eigenvalues of the plaquette terms violated at the boundary of $\mathcal S'$ are given by $\bs B_{p}^{3d} = -1$ and $\bs B_{p_{xy}} = \pm i$. Squaring the operator that creates the loop, we find that the only terms that are violated are $\bs B_{p_{xy}}=-1$, which correspond to $m^2$ excitations created at the positions where the original $m$ loop pierces each $xy$ plane. The $m^2$ excitations are planons because there are no local operators that can move them out of the plane.

Finally, moving the $e$ planon around the $m$ loop. we see that there is a single overlap of the $\mathcal Z$ and $\mathcal X$ operators, which results in a braiding phase of $i$.

In Appendix \ref{app:1foliatedGSD}, we calculate the ground state degeneracy of the model on a torus to be $2^{2L+3}$ and explicitly construct the logical operators.

\section{Fractonic Hybrid X-Cube Order}\label{3foliatedZ4Z2}

We now present a hybrid order that combines the phenomenology of a (three-foliated) fracton order with that of the 3d toric code.  Here, we will find that the charges and fluxes that were originally planons in the hybrid toric code layers will become fractons and lineons that behave similarly to the excitations of the X-Cube model. Furthermore, the fracton will square to a mobile topological charge, and thus we will refer to the resulting order as the fractonic hybrid X-Cube order. This hybrid order can be intuitively understood as the deconfined phase of a gauge theory, which is obtained by gauging a collection of symmetries -- including both global symmetries, as well as symmetries along three intersecting planes (planar subsystem symmetries) -- of a short-range-entangled phase.  

\subsection{Paramagnet with Global and Subsystem Symmetries}
To illustrate the gauging procedure, we  consider a four-dimensional Hilbert space at each vertex of a cubic lattice, with the $\ZZ_4$ clock and shift operators defined at each lattice site, as in Eq. \eqref{equ:clockshift}.
We may consider a product state with $\cX_{v} = +1$ at all lattice sites, which is trivially the ground-state of a Hamiltonian 
\begin{align}
    H =& -\sum_v [\cX_v+ \cX_v^2 + \cX_v^3].
\end{align}
We will consider the gapped symmetric excitations of this paramagnet which are invariant under a global $\ZZ_4$ symmetry transformation $\prod_v \cX_v$, along with a planar $\ZZ_2$ symmetry along any plane $\text{p}$ in the $xy$, $yz$ or $xz$ directions $\prod_{v \in \text{p}} \cX^2_v$. After gauging these symmetries, these excitations are in one-to-one correspondence with the gapped, \emph{fractionalized} charge excitations of the resulting hybrid phase.

 The elementary excitations of the paramagnet are created as follows.  First, for a plaquette $p$, the operator $\Delta_p =\cZ_i\cZ_j^\dagger\cZ_k\cZ_l^\dagger$ excites four charge excitations at the corners $i,j,k,l$. These excitatons are charged $\pm i$ under the global $\ZZ_4$ symmetry. In addition, they are also charged $-1$ under the $\ZZ_2$ planar symmetry, which renders them immobile. These excitations will correspond to the fracton $e$ after the gauging procedure. However, applying this operator twice creates four particles which are charged $-1$ under the global $\ZZ_4$, but charge neutral under the planar $\ZZ_2$. Therefore, these charges are mobile, and can be hopped using $\Delta_e = \cZ_i^2 \cZ_f^2$, where $i$ and $f$ are the endpoints of the edge $e$. Explicitly,
\begin{align}
    \Delta_{(ijkl)}^2 = \Delta_{(ij)}\Delta_{(ik)}\Delta_{(il)}.
    \label{eq:DeltahopZ4Z2}
\end{align}

\subsection{Hybrid Order}

We will now gauge the aforementioned symmetry, the details of which we will relegate to Appendix \ref{app:Z4Z2derivation}. Qualitatively, we separate the gauging into two steps. First, we gauge the $\ZZ_2$ planar symmetries. This results in a $\ZZ_2$ X-Cube model where the remaining $\ZZ_2$ global symmetry fractionalizes on the fracton excitation. Further gauging this $\ZZ_2$ global symmetry will give the hybrid model we will now present. Instead, we opt to motivate the resulting Hamiltonian as a hybridization between the 3d toric code and the X-Cube model.

\begin{figure}[t]
    \centering
    \includegraphics[scale=0.8]{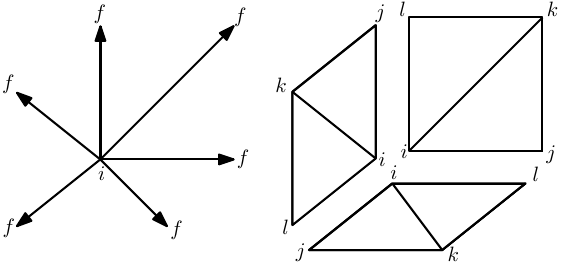}
    \caption{{\bf Description of the Lattice for the Hybrid X-Cube Model:} Diagonal edges are added to each plaquette in the cubic lattice. \emph{Left}: each edge $e=(if)$ is oriented, pointing outward from an ``initial" vertex $i$ towards a ``final" vertex $f$. \emph{Right}: ordering of vertices for each square plaquette $p=(ijkl)$}
    \label{fig:ordering}
\end{figure}

We consider a cubic lattice with an additional diagonal edge added to each plaquette on the cubic lattice as shown in Figure \ref{fig:ordering}, and place a $\ZZ_2$ gauge field (qubit) on each edge and each (square) plaquette of this lattice. In addition, we assign a local ordering of the vertices to each edge $e= (if)$ and each square plaquette $p=(ijkl)$ as shown in Fig. \ref{fig:ordering}.

\begin{figure*}[t!]
    \centering
    \begin{align*}
        \bs A_v &= \raisebox{-.5\height}{\includegraphics[scale=0.4]{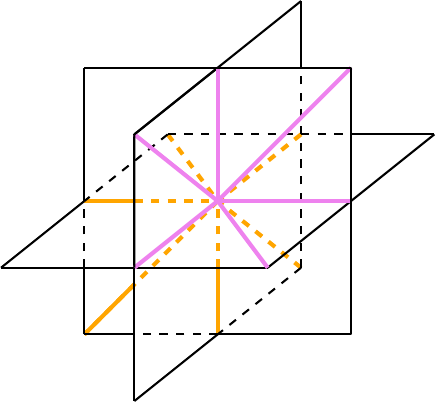}} & \bs A_v^2=\bs A_v^{XC} &= \raisebox{-.5\height}{\includegraphics[scale=0.4]{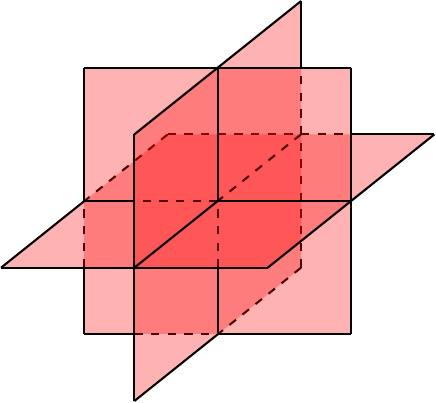}} & \bs B_{\nablapic} = \raisebox{-.5\height}{\includegraphics[scale=0.4]{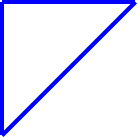}}
        \end{align*}
        \begin{align*}
         \bs B_{c,x}&= \raisebox{-.5\height}{\includegraphics[scale=0.4]{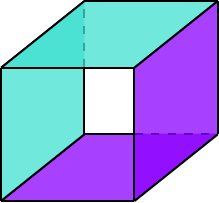}} &
         \bs B_{c,y}&= \raisebox{-.5\height}{\includegraphics[scale=0.4]{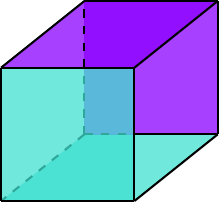}}
         &  \bs B_{c,z}&= \raisebox{-.5\height}{\includegraphics[scale=0.4]{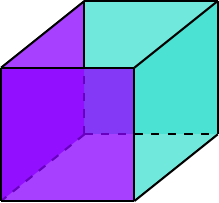}}\\
         \bs B_{c,x}^2 &= \raisebox{-.5\height}{\includegraphics[scale=0.45]{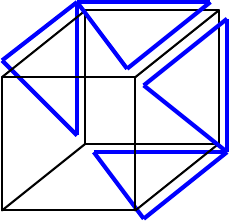}} &
         \bs B_{c,y}^2&= \raisebox{-.5\height}{\includegraphics[scale=0.45]{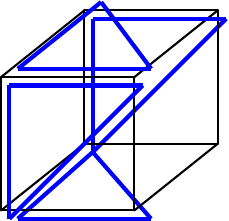}}
         &  \bs B_{c,z}^2&= \raisebox{-.5\height}{\includegraphics[scale=0.45]{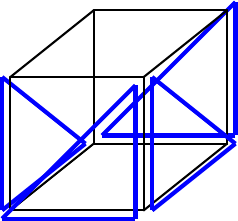}}
         \end{align*}
         \begin{align*}
         \bs \xi_e &=\begin{cases}
         \raisebox{-.5\height}{\includegraphics[scale=0.4]{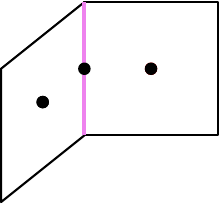}} \vspace{5pt}\\
         \raisebox{-.5\height}{\includegraphics[scale=0.4]{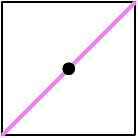}} 
         \end{cases}
         \equiv
         \begin{cases}
         \raisebox{-.5\height}{\includegraphics[scale=0.4]{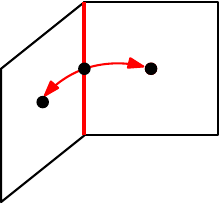}}\vspace{5pt}\\
         \raisebox{-.5\height}{\includegraphics[scale=0.4]{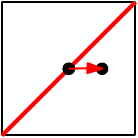}}
         \end{cases} &\bs \xi_e^2 &=
         \begin{cases}
         \raisebox{-.5\height}{\includegraphics[scale=0.4]{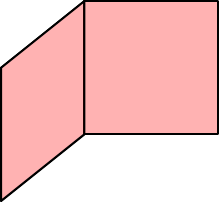}}\vspace{5pt}\\
         \raisebox{-.5\height}{\includegraphics[scale=0.4]{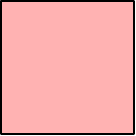}}
        \end{cases} &          \bs \zeta_p &= \raisebox{-.5\height}{\includegraphics[scale=0.4]{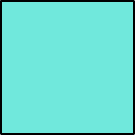}} \equiv \raisebox{-.5\height}{\includegraphics[scale=0.4]{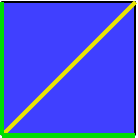}} & \bs \zeta_p^2 &= \raisebox{-.5\height}{\includegraphics[scale=0.4]{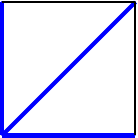}}
    \end{align*}
    \caption{{\bf Lattice Model for the Fractonic Hybrid X-Cube Order:} Visualization of the operators in the fractonic hybrid X-Cube model. The color coding used is red=$X$ , blue=$Z$, green = $S$, yellow = $S^\dagger$, magenta = $\bs \xi$, orange = $\bs \xi^\dagger$, cyan = $\bs \zeta$, purple = $\bs \zeta^\dagger$,{\color{red} $\rightarrow$}= \textsc{CNOT}.}
    \label{fig:Z4Z2}
\end{figure*}

The Hamiltonian can be thought of as first starting with a $\ZZ_2$ toric code defined with $\ZZ_2$ gauge fields on each edge of the lattice tensored with a $\ZZ_2$ X-Cube model defined with $\ZZ_2$ gauge fields on each square plaquette. Then, we couple the two gauge fields by modifying the vertex term in the toric code and the cube term in the X-Cube model. The Hamiltonian is given by  

\begin{align}
 H_\text{Hybrid} =& H_{TC}' + H_{XC}', \nonumber \\
     H_{TC}' =&-\sum_v \frac{\bs A_v+\bs A_v^\dagger}{2}  -  \sum_{\nablapic} \frac{1+\bs B_{\nablapic}}{2}, \nonumber\\
     H_{XC}' =& -\sum_v\frac{1+\bs A_v^{XC}}{2} - \sum_{c}\sum_{r=x,y,z} \frac{\bs B_{c,r} + \bs B_{c,r}^\dagger}{2},
     \label{equ:Z4Z2gaugetheorysimplified}
\end{align}
where
\begin{align}
\bs A_v &=\prod_{e \rightarrow v}\bs \xi_e^\dagger \prod_{e \leftarrow v}\bs \xi_e,\label{equ:Av} \\ 
\bs B_{{\nablapic}} &=\prod_{e \in {\nablapic}} Z_e ,
\label{equ:Bnabla}\\
\bs A_v^{XC} &=\prod_{p \supset v} X_p ,\label{equ:Av2}\\
\bs B_{c,r}&= \prod_{p \in c_r' }\bs \zeta_p^\dagger  \prod_{p \in c_r }\bs \zeta_p.
\label{equ:Bcr}
\end{align}
Here, $Z_e$ is the Pauli-$Z$ operator on each edge, and $X_p$ is the Pauli-$X$ operator on each plaquette. Visually, the operators above are shown in Fig. \ref{fig:Z4Z2}. To clarify the notation above, $e \rightarrow v$ ($e \leftarrow v$) in the vertex term $\bs A_v$ denotes the incoming (outgoing) edges towards (from) the vertex $v$ as defined in Fig. \ref{fig:ordering}, and shown in orange (magenta) in Fig. \ref{fig:Z4Z2}. For the plaquette term $\bs B_{\nablapic}$, the sum is over all triangular plaquettes ${\nablapic}$. The cube term $\bs B_{c,r}$, as in the X-Cube model, depends on an orientation $r$. In particular, $c_r$ and $c_r'$ for $r=x,y,z$, are each a set of two plaquettes surrounding the cube $c$ shown in cyan and purple respectively in Fig. \ref{fig:Z4Z2}. As in the X-Cube model, they satisfy $\bs B_{c,x} \bs B_{c,y} \bs B_{c,z} = 1$.

Now, we notice that if $\bs \xi_e$ and $\bs \zeta_p$ were Pauli-$X$ operators on each edge and Pauli-$Z$ operators on each plaquette, then this Hamiltonian is indeed just a stack of the toric code and the X-Cube models. However, in the hybrid model, the operators $\bs \xi_e$ and $\bs \zeta_p$ are Pauli operators decorated with additional Clifford gates

\begin{align}
    X_e\rightarrow \bs \xi_e &= X_e \prod_{p\in n(e)} \textsc{CNOT}_{e,p},
\label{equ:xi_e}\\
Z_p\rightarrow \bs \zeta_p&=Z_{p} S_{(ij)} S^\dagger_{(ik)}S_{(il)},\label{equ:zeta_p}
\end{align}
where
\begin{align}
   S= 
\begin{pmatrix}
1 & 0 \\
0 & i 
\end{pmatrix},~~~\textsc{CNOT} = 
\begin{pmatrix}
1 & 0 & 0 & 0 \\
0 & 1 & 0 & 0 \\
0 & 0 & 0 & 1 \\
0 & 0 & 1 & 0 \\
\end{pmatrix}.
\end{align}
Here, for the CNOT gate in Eq. \eqref{equ:xi_e}, the qubit on the edge $e$ is the control and the plaquette $p$ is the target. Furthermore, $n(e)$ is the set of plaquettes $p$ such that the edge $e$ appears as $(ij)$, $(ik)$, or $(il)$ of $p$ as defined in Fig.  \ref{fig:ordering} (See also Eq. \eqref{equ:Z4totwoZ2})

Because the operators $\bs \xi_e$ and $\bs \zeta_p$ act on both the gauge fields on edges and on plaquettes, we can anticipate that the excitations created by them will display features pertinent to both the toric code and  and X-Cube models. For example, as we will see, a fracton excitation can have non-trivial statistics with the flux loop. Let us analyze the algebra of these operators.

First, when restricted to terms only on plaquettes or on edges, the operators act the same as $\ZZ_2$ Pauli operators,
\begin{align}
    \bs \zeta_p X_{p'} &= (-1)^{\delta_{p,p'}}  X_{p'}\bs \zeta_p,\\
    Z_e \bs \xi_{e'} &=  (-1)^{\delta_{e,e'}}  \bs \xi_{e'} Z_e,\\
    [\bs \xi_e, \bs \xi_{e'}] = [\bs \zeta_p, \bs \zeta_{p'}]&=[\bs \xi_e, X_p] = [\bs \zeta_p, Z_e]=0.
\end{align}
Second, the square of the modified operators are also Pauli operators,
\begin{align}
    \bs \xi_e^2=\prod_{p\in n(e)}X_p,~~~\bs \zeta_p^2=Z_{(ij)}Z_{(ik)}Z_{(il)}.
\end{align}
Third, the operators $\bs \zeta_p$ and $\bs \xi_e$ either commute, or act as the clock and shift operators of a $\ZZ_4$ qudit, depending on if $e$ is a certain edge of the plaquette $p$ shown in Fig. \ref{fig:ordering},
\begin{align}
    \bs \zeta_p \bs \xi_e =\begin{cases}
    +i \bs \xi_e \bs \zeta_p;  & e= i_pj_p,i_pl_p, \\
    -i\bs \xi_e \bs \zeta_p; & e = i_pk_p,\\
    \ \ \ \ \ \bs \xi_e \bs \zeta_p; & \text{otherwise}.
    \end{cases} 
    \label{equ:xiezetapalg}
\end{align}
The first and second properties implies that $\bs \zeta_p$ is still a $\ZZ_2$ gauge field on plaquettes with respect to the electric field $X_p$ . Similarly, $\bs \xi_e$ is still the $\ZZ_2$ electric field on each edge with respect to the gauge field $Z_e$.  The replacement only modifies the star term $\bs A_v$ and the cage term $\bs B_{c,r}$. Because of the second property, the vertex term of the toric code $\bs A_v$ term squares to the vertex term of the X-Cube model $\bs A_v^{XC}$, and the cube term of the X-Cube model $\bs B_{c,r}$ squares to a product of four triangular plaquette terms $\bs B_{\nablapic}$ of the toric code as shown in Fig. \ref{fig:Z4Z2}.

\subsubsection{Excitations and fusion}
\begin{table*}[t!]
\centering
\caption{\textbf{Excitations in the fractonic hybrid X-Cube model:} A summary of the pure charge and flux excitations in the fractonic hybrid X-Cube model is provided above, along with the local operators that measure these excitations in the lattice model.}
\def\arraystretch{1.5}
\begin{tabular}{ | l  | l |l| l|}
\hline
\multicolumn{1}{|c|}{Excitation} & \multicolumn{1}{c|}{Creation operator} &  \multicolumn{1}{c|}{Charges} & \multicolumn{1}{c|}{Local Wilson operator}\\
\hline
fracton $e$ & Corners of $\bs \zeta_p$ membrane& $\bs A_v = i$, $\bs A_v^{XC}=-1$ &$\bs B_{c}=$  Closed loop of $e-\bar e$ dipole around $c$\\
\hline
\multirow{2}{*}{mobile charge $e^2$} &  Corners of $\bs \zeta_p^2$ membrane &\multirow{2}{*}{$\bs A_v = -1 $}&\multirow{2}{*}{$\bs B_{\nablapic}=$ Closed loop of $e^2$ around $\nablapic$.}\\
& End points of $Z_e$ string&&\\
\hline
loop $m$ &  Boundary of $\bs \xi_e$ membrane &$\bs B_{p}= \pm i$ (at corners), $ \bs B_{\nablapic}= -1$ &$\bs A_v=$ Closed membrane of $m$ around $v$\\
\hline
\multirow{2}{*}{lineon $m^2$} & Corners of $\bs \xi_e^2$ membrane & \multirow{2}{*}{$\bs B_{p}= -1$} &\multirow{2}{*}{$\bs A_v^{XC}=$ Closed cage of $m^2$ around $v$}\\
& End point of $X_p$ string &&\\
\hline
\end{tabular}
\label{tab:exc_3foliated}
\end{table*}
Since the model is a commuting projector Hamiltonian, it is exactly solvable. Therefore, we can explicitly write down the excitations and compare the similarities to the hybrid toric code layers in the previous section. Using the commutation relations Eq. \eqref{equ:xiezetapalg}, we see that $\bs \zeta_p$ commutes with $\bs B_{c,r}$ and $\bs B_{\nablapic}$, but violates the projector containing $\bs A_v$ at the four vertices at the corners of $p$. In particular, this implies that the four corners of $\bs \zeta_p$ are charged $\pm i$ under the operator $\bs A_v$. Furthermore, since $\bs A_v^2 = \bs A_v^{XC}$ these excitations are also charged $-1$ under $\bs A_v^{XC}$ and are therefore fractons. We will call the excitations $e, \bar e$ for the excitation $\bs A_v = i,-i$, respectively. In general, a product of $\bs \zeta_p$ over a surface $\mathcal S$ creates such fractons at the corners of $\mathcal S$
\begin{align}
    \text{fracton}~e,\bar e:\prod_{p\in \mathcal S}\bs \zeta_p.\label{eq:e}
\end{align}
The other type of charge excitation is created by a product of $Z_e$ on an open string $L$,
\begin{align}
    \text{mobile}~e^2:~~~\prod_{e\in L} Z_e. \label{eq:e2string}
\end{align}
The end points of the string operator above are charged $-1$ under $\bs A_v$, and commute with other local terms in the Hamiltonian.  Now, the operator $\bs \zeta_p^2$ also creates such excitations on the corners of $p$, as it is charged $-1$ under $\bs A_v$ at the corners. Thus, we will call $\bs A_v=-1$ the point excitation $e^2$.  This can be seen from the fact that $\bs \zeta_p^2$ can be written as a product of $Z_e$ operators. More generally, the product of $\bs \zeta_p^2$ on a surface $\mathcal S$ creates the point charges at the corners of $\mathcal S$,
\begin{align}
    \text{mobile}~e^2 \text{ on corners}:\prod_{p\in \mathcal S} \bs \zeta_p^2,~~~\bs \zeta_p^2 = Z_{(ij)} Z_{(ik)} Z_{(il)},\label{eq:e2}
\end{align}
which is just the dualized form of Eq. \eqref{eq:DeltahopZ4Z2}.

The $e^2$ excitations are fully mobile, since a string of $Z_e$ operators can hop individual $e^2$ excitations. This can also be seen from the fact that it is not charged under $\bs A_v^{XC}$, which detects the fracton.

\begin{figure}[t]
    \centering
    \includegraphics[scale=0.45]{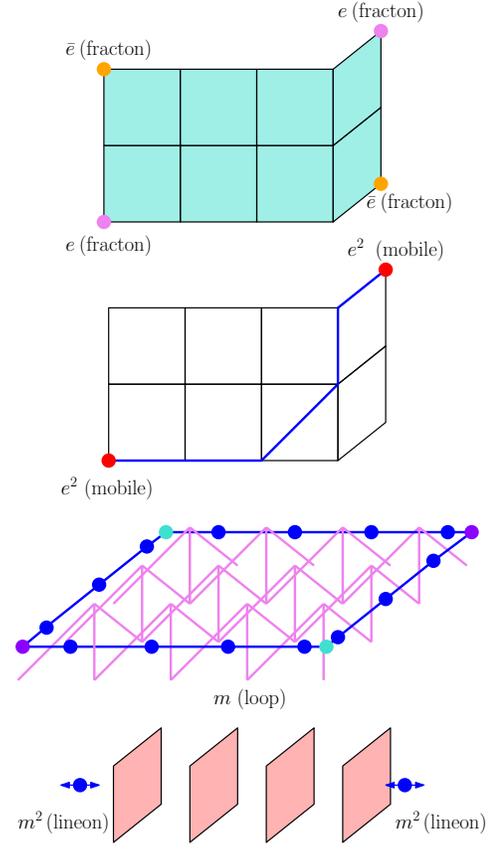}
    \caption{{\bf Geometry of the Excitations in the Hybrid X-Cube Order:} Excitations of the fractonic hybrid X-Cube model and their corresponding creation operators are shown. The excitations $e$, $e^2$, $m$ and $m^2$ are created using a membrane of $\bs \zeta_p$ (cyan), a flexible string of $Z_e$ (blue), a membrane of $\bs \xi_e$ (magenta), and a rigid string of $X_p$ (red), respectively.}
    \label{fig:exc_3foliated}
\end{figure}

Next, we define operators that violate $\bs B_p$ and $\bs B_{\nablapic}$, but commute with $\bs A_v$ and $\bs A_v^{XC}$. Excitations created from such operators are flux excitations. First, acting with $\bs \xi_e$ on all edges intersecting a given surface $\mathcal S'$ on the dual lattice creates a loop excitation at the boundary of that surface.
\begin{align}
    \text{mobile loop}~m:~~~\prod_{e\perp \mathcal S'} \bs \xi_e, \label{eq:mloop}
\end{align}
More precisely, the operators $\bs B_{\nablapic}$ along the boundary of $\mathcal S'$ are charged $-1$. Interestingly, we find that the corners of the loop operators are moreover charged $\pm i$ under two of the three $\bs B_c$ operators. This is shown in cyan and purple in Fig, \ref{fig:exc_3foliated}.

Lastly, applying $X_p$ violates $\bs B_{c,r}$ on the two cubes adjacent to $p$, creating two lineon excitations. In general, the lineon operator is the product of $X_p$ along a rigid string $L'$ on the dual cubic lattice,
\begin{align}
    \text{lineon}~m^2:~~~\prod_{p\perp L'} X_p,
\end{align}
and are charged $-1$ under two of the three $\bs B_{c,r}$ operators. In particular, a lineon mobile in the direction $r'=x,y,z$ is charged under the $\bs B_{c,r}$ for $r' \ne r$.

The excitations of this model are summarized in Table \ref{tab:exc_3foliated} and shown in Figure \ref{fig:exc_3foliated}.

\begin{figure*}[t]
    \centering
    \includegraphics[scale=0.52]{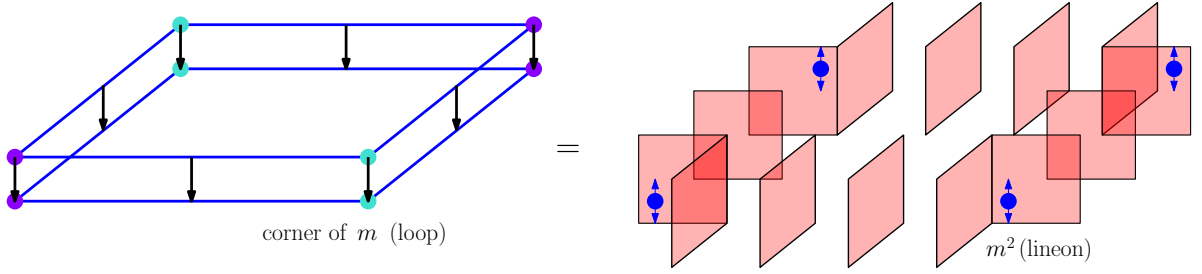}
    \caption{{\bf Loop fusion in the fractonic hybrid X-Cube order:} Fusion of two identical $m$ excitations results in lineon excitations ($m^2$) at the corners of the membrane. The direction of mobility (blue double arrow) for each lineon is perpendicular to the two segments of the loop meeting at that corner.}
    \label{fig:mloopfusion}
\end{figure*}

These excitations also have interesting fusion rules. Two fractons fuse into a mobile charge, as we can see from comparing Eq. \eqref{eq:e} with Eq. \eqref{eq:e2}. More surprisingly, two identical loop excitations fuse into a number of lineons. To see this, we consider a fusion of the loop $m$ with itself by applying
\begin{align}
    \bs \xi_e^2 &= \prod_{p\in n(p)} X_p,
\end{align}
to all edges in a dual surface $\mathcal S'$. One can verify that
\begin{align}
    \prod_{e \in \mathcal S'}\bs \xi_e^2 &= \prod_{p\in \partial \mathcal S'} X_p.
\end{align}
That is, the product is equal to applying $X_p$ to all plaquettes along the boundary of $\mathcal S'$, which is just the lineon string along the original loop. Therefore, a lineon excitation is created at every corner of the original loop excitation, and the lineon is mobile along the direction normal to both edges of the loop meeting at the given corner. This is illustrated in Figure \ref{fig:mloopfusion}. 

To summarize our results, the excitations in this exactly solvable model can be created by cutting open ``closed" Wilson operators. The excitations are ``topological" in the sense that once created, they can fluctuate as far as their mobilities allow without an energy cost. To see this, we point out that the following Wilson operators are just products of the stabilizers, and therefore commute with the Hamiltonian 
\begin{align}
W^{e^2}(L)=&\prod_{e\in L} Z_e=\prod_{\nablapic\in S} \bs B_{\nablapic},\\
W^{m}(\Sigma)=&\prod_{e\perp \Sigma} \bs \xi_e =\prod_{v\in \mathcal V} \bs A_v, \\
W^{e-\bar e}_r (S_r)=&\prod_{p\in S_r} \bs \zeta_p = \prod_{c\in \Omega_r } \bs B_{c,r}, \\
W^{m^2 ~\text{cage}}(C)=& \prod_{p \in C} X_p =\prod_{v\in V} \bs A_v^{XC}.
\end{align}
Here, $L$ is a loop which encloses a surface $S$, $\Sigma$ is a surface which encloses a volume $\mathcal V$, $S_r$ is a closed ribbon within a plane perpendicular to $\hat r$ which encircles the region $\Omega_r$, and $C$ is a rigid cage configuration which encloses a block volume $V$. This means that open Wilson operators that act on different sub-manifolds but share the same boundary will create the same excitations.  For example, the $m$ loop excitation, though created by a surface operator, does not depend on the choice of surface in which we choose to fill the loop.

\subsubsection{Braiding}
\begin{table}[b]
\centering
\caption{{\bf Braiding Data:} Summary of the braiding phases in the fractonic hybrid X-Cube model. The braiding process is obtained by applying the closed Wilson operator over the excitations of the fractonic hybrid X-Cube model, as shown in Figs. \ref{fig:Haah_Hybrid}e, and \ref{fig:embraidcombined}.}
\def\arraystretch{1.5}
\begin{tabular}{ | c  |c |}
\hline
 &  \makecell{Closed membrane \\ of $m$ $(W^m(\Sigma))$} \\
 \hline
fracton $e$ &  $i$\\
charge $e^2$ &  $-1$ \\
\hline
\end{tabular}
\vspace{5mm}

\begin{tabular}{ | c  |c c |}
\hline
 & \makecell{Closed string of \\ $e-\bar e$ dipole $(W_r^{e-\bar e}(S_r))$} & \makecell{Closed string of  \\ $e^2$ charge $(W^{e^2}(C))$}  \\
 \hline
loop $m$ & \makecell{$\pm i$\\ {\small (at corners of the loop)}} & $-1$  \\
lineon $m^2$ & $-1$ & $1$  \\
 \hline
\end{tabular}
\label{tab:braiding}
\end{table}
The  hybrid model has non-trivial braiding processes between the excitations of different mobilities. Similar to usual topological phases, we can prepare an excitation using an open Wilson operator, then act with a different closed Wilson operator to perform the braiding. The closed Wilson operator describes the limit of a braiding process in the space-time picture, performed in an infinitesimally small amount of time. 

\begin{figure}
     \centering
     \includegraphics[scale=0.56]{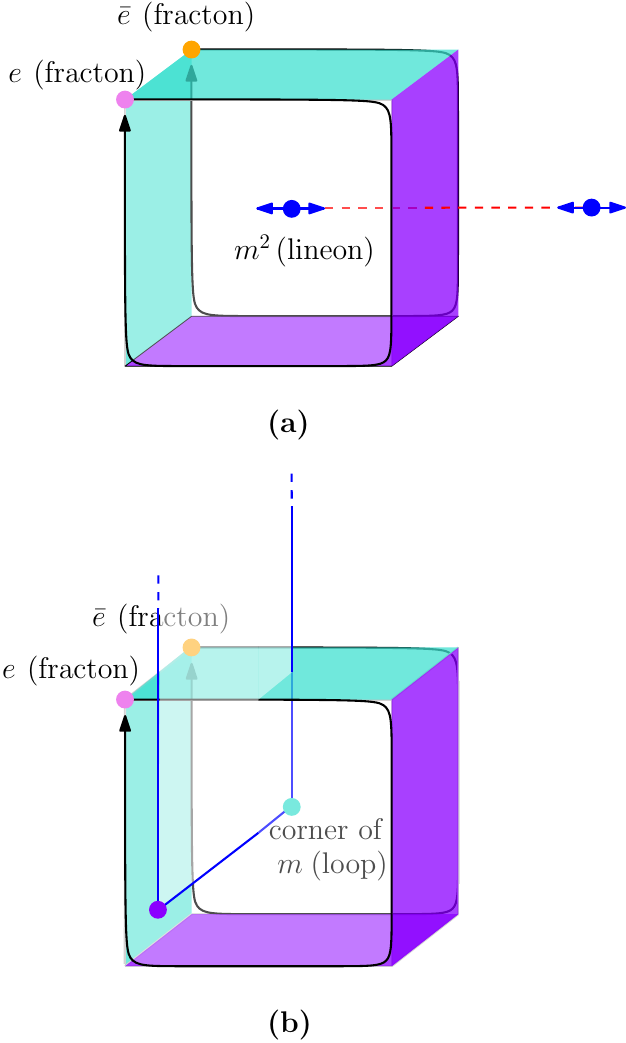}
     \caption{{\bf Braiding Processes for an $\bs e \bs -\bar{\bs e}$ Dipole:} \textbf{(a)} The braiding of an $e-\bar e$ dipole (pointing in the $x$ direction) with an $m^2$ lineon mobile along the $y$ direction. As the dipole moves around a closed loop in the $yz$ plane, if the dipole winds around the lineon, it picks up a phase of $-1$. \textbf{(b)} An analogous braiding of an $e-\bar e$ dipole with a corner of the $m$ loop in $xz$ plane. If the path of the $e$ fracton forms a link with the $m$ loop, it picks up a phase of $i$. The two braiding processes are consistent with the fusion of two $m$ loops in Fig. \ref{fig:mloopfusion}.}
     \label{fig:embraidcombined}
 \end{figure}

As in a $\ZZ_4$ gauge theory, there is a braiding phase of $-1$ between the mobile charge $e^2$ and the flux loop $m$, reminiscent of the braiding in a $\ZZ_2$ toric code. We show this in Fig. \ref{fig:Haah_Hybrid}e. Similarly, there is also a braiding phase of $-1$ between a fracton dipole pointing in direction $\hat r$ with a lineon mobile along the $\hat r'$ direction if $r \ne r'$ as shown in Fig. \ref{fig:embraidcombined}(a). \footnote{This braiding phase is well-defined regardless of how the lineon is created--whether from a single string operator, or from the fusion of two lineons mobile in the other two directions.} The braiding process here can be compared to the fracton-lineon braiding process in the X-Cube model\cite{SlagleKim2017,PaiHermele2019,BulmashIadecola19}.

The more interesting braiding that makes this model different from a stack of the toric code and X-Cube models is that the fracton $e$ and the flux loop $m$ has an Aharonov-Bohm phase of $i$. There are two ways to see this. One way is to realize that acting with the closed Wilson operator $W^m(\Sigma)$ around a fracton braids a flux loop around that vertex. Since $W^m(\Sigma)$ is just a product of $\bs A_v$ operators enclosed within the surface $\Sigma$, and the fracton is charged $i$ under $\bs A_v$, this implies that the braiding process gives a statistical phase of $i$.
 
 Alternatively, we propose an exotic braiding process between a fracton dipole and a corner of an $m$ loop, as shown in Figure \ref{fig:embraidcombined}(b). We consider a fracton dipole and use the closed Wilson operator $W_r^{e-\bar e}(S_r)$ to hop the fracton dipole in a closed trajectory perpendicular to the direction $r$. Since $W_r^{e-\bar e}(S_r)$ is a product of $\bs B_{c,r}$ operators, and the corner of the $m$ loop is charged $\pm i$ under $\bs B_{c,r}$ in two of the three directions, we find that if the trajectory of the fracton dipole encloses a corner of the $m$-loop within the same plane, then the process can detect a phase of $\pm i$. Specifically, there is a statistical phase if the $m$-loop pierces the $W_r^{e-\bar e}(S_r)$ membrane. It is interesting to note that although the fracton is immobile, it is allowed to move when paired up as a dipole. Furthermore, it is only when the dipole braids with a corner of the $m$ loop that only one of the fractons winds up forming a link with the $m$-loop\footnote{One might be concerned that the notion of a dipole detecting a corner of an $m$-loop might not be well-defined away from the exactly solvable limit. In particular, whether the notion of a $m$ loop corner is well defined point in space if the membrane operator that creates the loop excitation has a larger support. However, we know that in the exactly solvable limit, the Hamiltonian has a conservation law that the product of $\bs B_{c,r}$ on all cubes in a given plane perpendicular to the direction $r$ is the identity. Since $\bs B_{c,r}$ detects the corners of the flux loop, this conservation law guarantees that each plane always has an even number of flux loop corners. Therefore, the notion of a flux loop corner is well defined for every plane. It follows that away from the exactly solvable limit, there is an equivalent conservation law (adiabatically connected to the $\bs B_{c,r}$ operator) that pins the flux loop corners to specific planes. Hence, the braiding process is well-defined throughout the hybrid phase.}.

 To conclude, the loop excitation in this hybrid model has exotic braiding properties which makes it distinct from a loop excitation in a pure \TQFT. Though the loop is fully mobile, its corners can be detected with a phase $i$ by fracton dipoles defined in the same plane (in two of the three directions). Note that this braiding is also consistent with fusion, since the $m$ loop corners square to lineons, which can be detected with an identical process with braiding phase $-1$.

Finally, it is important to point out certain braiding processes with trivial statistics. The first is the trivial braiding between the mobile charge $e^2$ and the lineon $m^2$. This can be seen from the fact that the operators that excite each particle do not overlap (one acts on edges, while the other acts on plaquettes). Furthermore, since $e^2$ is mobile and both are point particles, any possible braiding is homotopic to a trivial braiding process. The second is trivial three-loop braiding statistics, and other non-Abelian braiding processes which are important topological invariants for (liquid) 3d topological orders \cite{WangLevin2014,JianQi14,JiangMesarosRan2014,WangLevin2015,WangWen2015,PutrovWangYau2017,ChengTantivasadakarnWang2018,WangChengWangGu19,ChanYeRyu18,ZhouWangWangGu19,ZhangYe20}. To show this, we use the fact that the $m$-loop can be excited by a membrane of $\bs \xi_e$, which satisfies $[\bs \xi_e, \bs \xi_{e'}]=0$. Therefore, any braiding of loops cannot produce a phase, including any three-loop braiding processes. In addition, because all operators in the algebra commute up to a phase, all braiding processes are Abelian.

A summary of the braiding phases are given in Table \ref{tab:braiding}.

\subsubsection{Ground state degeneracy and logical operators on a torus}\label{subsubsec:logicalop}
 
The ground state degeneracy of the hybrid model can be calculated via similar methods used for the toric code and X-Cube models. In Appendix \ref{app:GSD}, we count the number of independent stabilizers and compare it to the total dimension of the Hilbert space. We find that the ground state degeneracy of the model on a torus of size $L_x \times L_y \times L_z$, is 
\begin{align}
    \log_2 \textsc{GSD} = 2(L_x+L_y+L_z).
\end{align}

To distinguish the different ground states, we restrict ourselves to the ground state subspace and explicitly construct the logical (non-local Wilson) operators in this subspace by tunneling excitations around the torus. In the following, we argue that the logical operators can be factored to a ``toric code" subspace, consisting of operators that tunnel $e^2$ and $m$-loops, and an ``X-Cube" subspace, consisting of operators that tunnel $e$-dipoles and $m^2$-lineons. This will allow us to conclude that the ground state degeneracy is 
\begin{align}
     \log_2 \textsc{GSD} &= \log_2 \textsc{GSD}_{TC} + \log_2 \textsc{GSD}_{XC} \\
     &=3 + 2(L_x+L_y+L_z)-3 = 2(L_x+L_y+L_z).\nonumber
\end{align}
Therefore, they form a complete set of logical operators.

\begin{figure}[t!]
    \centering
    \includegraphics[scale=0.15]{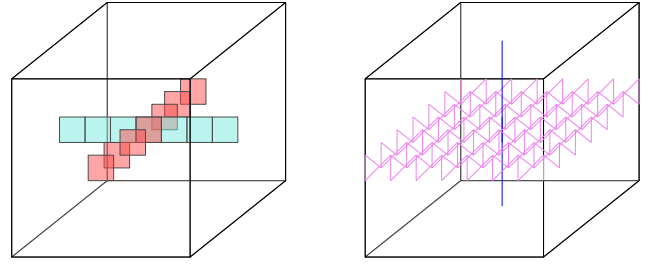}
    \caption{ {\bf Logical (non-local Wilson loop) operators of the Hamiltonian \eqref{equ:Z4Z2gaugetheory} on a three-torus:}  Left: $W^{e-\bar e}(R)$ and $W^{m^2}(L')$  logical operators of the ``X-Cube" subspace. They describe the tunneling of $e-\bar e$ fracton dipole (cyan) and $m^2$ lineons (red), respectively. Right: $W^{e^2}(C)$ and $W^{m}(\Sigma')$ logical operators of the ``toric code" subspace. They describe the tunneling of $e^2$ mobile charge (blue) and $m$ mobile loop (magenta), respectively.}
    \label{fig:logicalops}
\end{figure}

First, consider tunneling an $e-\bar e$ dipole around the torus using the operator
\begin{align}
    W^{e-\bar e}(R)=&\prod_{p\in R}\bs \zeta_p,
\end{align}
where $R$ is a cyan ribbon shown in Fig. \ref{fig:logicalops}. Now, although the corners of $\bs \zeta_p$ are fractons with $\ZZ_4$ fusion rules, the ribbon of $\bs \zeta_p$ is actually a $\ZZ_2$ operator in the ground state subspace, since it squares to a product of $\bs B_{\nablapic}$ operators, which is set to one. This operator anticommutes with the following operator that tunnels the $m^2$ lineon
\begin{align}
    W^{m^2}(L')=&\prod_{p\perp L'}X_p,
\end{align}
for some rigid string $L'$ that intersects the ribbon $R$ (shown in red). These set of operators form $2(L_x+L_y+L_z)-3$ pairs of independent $\ZZ_2$ logical operators identically to those in the X-Cube model.

Next, consider tunneling the $m$-loop around a non-trivial 2-cycle $\Sigma'$ of the torus (shown in magenta), which can be implemented by applying
\begin{align}
   W^{m}(\Sigma')=&\prod_{e\perp \Sigma'}\bs \xi_e.
\end{align}
Similarly, this operator is a $\ZZ_2$ operator in the ground state subspace, since squaring this operator gives at most a product of $\bs A_v^{XC}$ operators. This operator anticommutes with
\begin{align}
    W^{e^2}(C)=&\prod_{e\in C}Z_e,
\end{align}
for some 1-cycle $C$ (shown in blue) that intersects transversally with $\Sigma'$. On a torus, there are $3$ such pairs.

Lastly, the pairs $W^{e-\bar e}$ and  $W^{m}$ commute, which can be argued from the fact that an $e$-dipole braids trivially with an $m$-loop when there are no corners\footnote{More explicitly, for each plaquette $p$ on which the two Wilson operators overlap, $W^m$ contains $\bs \zeta_p$ and $W^{e-\bar e}$ contains either $\bs \xi_{ik}\bs \xi_{ij}$ or $\bs \xi_{ik}\bs \xi_{il}$. These two sets of operators always commutes using Eq. \eqref{equ:xiezetapalg}.}. This ensures that the logical operators factor into the two subsets as claimed. 

 We point out that the hybrid X-cube model can also be defined on different spatial manifolds, with a proper choice of foliation structure. For example, we can put the hybrid model on the manifold $S^3$ by first beginning with the X-cube model defined on $S^3$ in Ref.  \onlinecite{ShirleySlagleWangChen2018}, and apply the operator replacement in Eqs. \eqref{equ:xi_e}-\eqref{equ:zeta_p}. We expect that by comparing the number of independent constraints and the total dimension of the Hilbert space (as outlined in App. \ref{app:generalGSD}), one finds a unique ground state. Thus, the hybrid model is topologically ordered, and its ground state degeneracy depends on the topology of the spatial manifold.

\section{Lineonic Hybrid X-Cube Order }\label{sec:Z42Z22lineon}
\begin{table*}[t!]
\centering
\caption{\textbf{Excitations in the lineonic hybrid X-Cube model:} A summary of the pure charge and flux excitations in the lineonic hybrid X-Cube model is provided above, along with the local operators that measure these excitations in the lattice model.}
\def\arraystretch{1.5}
\begin{tabular}{ | l  | l |l| l|}
\hline
\multicolumn{1}{|c|}{Excitation} & \multicolumn{1}{c|}{Creation operator} &  \multicolumn{1}{c|}{Charges} & \multicolumn{1}{c|}{Local Wilson operator}\\
\hline
lineon $e_x$ & End point of $\cZ$ on $x$ edges & $\bs A_{v} = i$, $\bs A_{v,y}^{XC}=\bs A_{v,z}^{XC} = -1$ &\multirow{3}{*}{$\bs B_{c}=$  Cage of $e_x,e_y,e_z$ around $c$}\\
lineon $e_y$ & End point of $\cZ$ on $y$ edges & $\bs A_{v} = i$, $\bs A_{v,x}^{XC}=\bs A_{v,z}^{XC} = -1$  &\\
lineon $e_z$ & End point of $ZI$ on $z$ edges &  $\bs A_{v,x}^{XC}=\bs A_{v,y}^{XC} = -1$  &\\
\hline
\multirow{2}{*}{mobile charge $e^2$} &  End point of $\cZ^2$ on $x,y$ edge, &\multirow{2}{*}{$\bs A_{v} = -1 $}&\multirow{2}{*}{$\bs B_{p}=$ Closed loop of $e^2$ around $p$.}\\
& End points of $IZ$ on $z$ edge&&\\
\hline
\multirow{2}{*}{loop $m$} &  Boundary of $IX$ membrane in $xy$ plane &$ \bs B_{p}= -1$ &\multirow{2}{*}{$\bs A_{v}=$ Closed membrane of $m$ around $v$}\\
& Boundary of $\cX$ membrane in $xz,yz$ plane & $ \bs B_{p}= -1$, $\bs B_c = \pm i$ (at corners) &\\
\hline
\multirow{2}{*}{fracton $m^2$} & Corners of $XI$ membrane in $xy$ plane &\multirow{2}{*}{$\bs B_c = -1$} &\multirow{2}{*}{$\bs A_{v,r}^{XC}=$ Closed loop of $m^2$ dipole around $v$}\\
 &Corners of $\cX^2$ membrane in $xz,yz$ plane &&\\
\hline
\end{tabular}
\label{tab:exc_3foliatedlineon}
\end{table*}

In the previous model, the fractons and lineons were treated as charge and flux excitations respectively. We will now consider the opposite scenario, where the lineons are charges and the fractons are fluxes. The model in this section is therefore an example of a different type of hybridization between the toric code and the X-Cube model. To distinguish it from the former, we will refer to this hybridization as the Lineonic hybrid X-Cube model.

We remark that although this is the simplest model to construct in the case that lineons are charges, the model is anisotropic. As we will see, only lineons mobile in the $x$ or $y$ direction will square to a mobile particle, while the lineon mobile in the $z$ direction will square to the vacuum superselection sector. This is because the fusion rule of the three lineons $e_x \times e_y \times e_z =1$ forbids all three lineons from squaring to the same mobile $\ZZ_2$ particle. Nevertheless, it is possible to construct a different hybrid model where the lineons square to two different mobile particles. Such a model would instead be a hybrid between the X-Cube model and two 3d toric codes. We construct such a model explicitly in Appendix \ref{app:Z44Z22lineon}.

Following the structure of the previous section, the Ising model and its hybrid model are described in Secs. \ref{sec:Isinglineon} and \ref{sec:lineonham}, respectively. In Appendix \ref{sec:pstringlineon}, we show that this model can be obtained by a similar $p$-string condensation \cite{MaLakeChenHermele2017,Vijay2017} to the X-Cube model by replacing the stacks of toric codes in the $xy$ planes with the hybrid toric code layers of Sec. \ref{1foliatedZ4Z2}.

\subsection{Paramagnet with Global and Subsystem Symmetries }\label{sec:Isinglineon}
To obtain the previous model, the paramagnet had an onsite planar symmetry in three directions, where each onsite term is generated by the same normal subgroup of the global $\ZZ_4$ symmetry. Therefore, an excitation is charged under planar symmetries along all three directions, resulting in an immobile charge in the gauged model 

To start off differently, our paramagnet now has a global $G=\ZZ_4 \times \ZZ_2 = \inner{a^4=b^2=1}$ symmetry. However, the planar symmetry $N$ for each direction of planes is generated by a different subgroup of $G$. In particular, the $xz$, $yz$ and $xy$ planar symmetries are generated by the $\ZZ_2$ subgroups $a^2$, $a^2b$ and $b$, respectively. As a result, excitations of this paramagnet are charged under only two of the three planar symmetries and are therefore lineons. 

To obtain the model we are to present, we first gauge the planar symmetries of the model to obtain the X-Cube model. The remaining $\ZZ_2$ global symmetry fractionalizes on the lineon mobile in the $x$ and $y$ directions. We can then gauge the global $\ZZ_2$ symmetry to obtain the hybrid model.

\subsection{Hybrid Order}\label{sec:lineonham}

The model is defined on a cubic lattice with a $\ZZ_4$ qudit on each $x$ and $y$ edge, and two qubits on each $z$ edge. The Hamiltonian of the model is
\begin{align}
    H_\text{Hybrid} =& H_{TC}'+ H_{XC}', \nonumber\\
    H_{TC}' =& -\sum_v \frac{\bs A_{v}+\bs A_{v}^\dagger}{2} - \sum_{p} \frac{1 + \bs B_{p}}{2}, \nonumber\\
    H_{XC}' =& -\sum_v\sum_{r=x,y,z}  \frac{1+ \bs A_{v,r}^{XC}}{2} -\sum_c \frac{\bs B_c + \bs B_c^\dagger}{2},
    \label{equ:Hamlineonani}
\end{align}
where
\begin{align}
\bs A_{v} &=\raisebox{-0.5\height}{\includegraphics[scale=1]{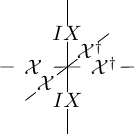}}, \\
\bs B_{p} &=\raisebox{-0.5\height}{\includegraphics[scale=1]{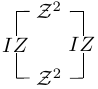}},~~~ \raisebox{-0.5\height}{\includegraphics[scale=1]{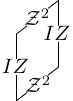}},~~~ \raisebox{-0.5\height}{\includegraphics[scale=1]{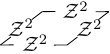}},
\end{align}
    \begin{align}
    \bs A_{v,x}^{XC} &=\raisebox{-0.5\height}{\includegraphics[scale=1]{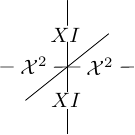}}, & \bs A_{v,y}^{XC} &=\raisebox{-0.5\height}{\includegraphics[scale=1]{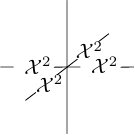}} , \\
    \bs A_{v,z}^{XC} &=\raisebox{-0.5\height}{\includegraphics[scale=1]{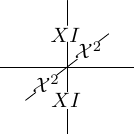}}, &  \bs B_{c} &=\raisebox{-0.5\height}{\includegraphics[scale=1]{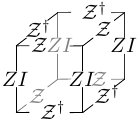}}.
    \end{align}
Note that the X-Cube model is defined on the dual cubic lattice compared to that of Sec. \ref{3foliatedZ4Z2}. The vertex terms of the X-Cube model satisfy, $\bs A_{v,x}^{XC} \bs A_{v,y}^{XC} \bs A_{v,z}^{XC}=1$. Furthermore, because of the hybridization, the vertex term of the toric code $\bs A_v$ squares to $\bs A_{v,z}^{XC}$ of the X-Cube model, and the cube term $\bs B_c$ of the X-Cube model squares to a product of two $\bs B_{p}$ plaquettes.

\subsubsection{Excitations, fusion, and braiding}

\begin{figure}[t!]
    \centering
    \includegraphics[scale=.9]{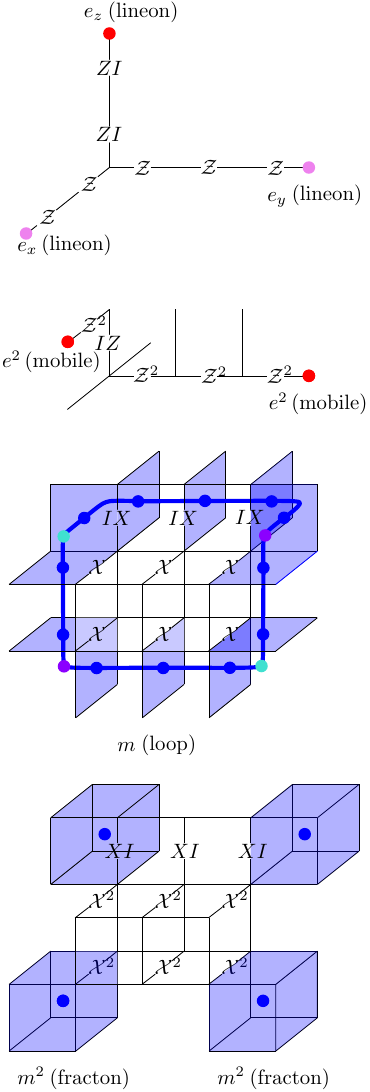}
    \caption{{\bf Excitations and their Creation Operators in the Lineonic X-Cube Model}}
    \label{fig:exc_lineon_ani}
\end{figure}
The excitations in this model are shown in Fig. \ref{fig:exc_lineon_ani}. First we discuss the charges, which are violations of the vertex terms. Because this model is anisotropic, the lineons $e_x$ and $e_y$ in the $xy$ plane are excited with $\cZ$ on a rigid string in the $xy$ plane. On the other hand, the lineon $e_z$ is excited with $ZI$ on a rigid string in the $z$ direction. The lineon $e_r$ correspond to $\bs A_{v,r'}^{XC}=-1$ for $r \ne r'$. Note that like the X-Cube model, the three lineons fuse to the vacuum.
\begin{align}
    \text{lineon}~ \begin{cases}
    e_x,\bar e_x :& \prod_{e \in L_x} \cZ_e\\
     e_y,\bar e_y :& \prod_{e \in L_y} \cZ_e\\
      e_z,\bar e_z :& \prod_{e \in L_z} ZI_e
    \end{cases}.
\end{align}
The fusion rules for each lineon species, however, is different. 
\begin{align}
    e_z \times e_z&=\text{ vacuum}\nonumber\\
    e_x \times e_x&= e_y\times e_y \equiv e^2.
\end{align} The excitation $e^2$ is a mobile particle, and can move in the $xy$ plane using $\cZ^2$, as well as in the $z$ direction using $IZ$.
\begin{align}
    \text{mobile}~ e^2: \prod_{e \in L_x,L_y} \cZ^2_e,\prod_{e \in L_z} IZ_e.
\end{align}
This mobile excitation is charged $-1$ under $\bs A_v$.

Next, we discuss the flux excitations. The first is the $m$-loop, which is a violation of plaquettes. To excite an $m$ loop, we apply $IX$ on a $z$-edge , or $\cX,\cX^\dagger$ on an $x$ or $y$ edge. However, we also notice that when the $m$-loop is oriented in the $xz$ or $yz$ plane, the corners of the $m$-loop are charged $\pm i$ under the $\bs B_c$ operator as shown in cyan and purple in Fig. \ref{fig:exc_lineon_ani} .
\begin{align}
    \text{loop}~ m: \prod_{e \perp \mathcal S_{xy}} IX_e ,  \prod_{e \perp \mathcal S_{xz},\mathcal S_{yz} } \cX_e.
\end{align}
Finally, the fracton is the excitation $\bs B_c = -1$. Four fractons can be created on the four cubes adjacent to an edge using $\cX^2$ acting on an $x$ or $y$ edge, or using $XI$ on a $z$ edge.
\begin{align}
    \text{fracton}~ m: \prod_{e \perp \mathcal S_{xy}} XI_e ,  \prod_{e \perp \mathcal S_{xz},\mathcal S_{yz} } \cX^2_e.
\end{align}

\begin{figure}
    \centering
    \includegraphics[scale=0.37]{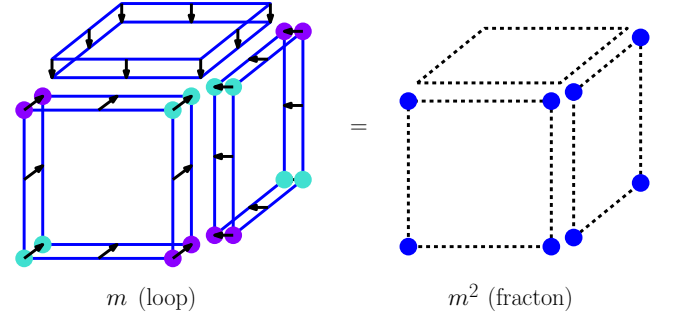}
    \caption{{\bf Loop fusion in the lineonic hybrid X-Cube order:} After fusing two identical $m$ loops in the lineonic hybrid X-Cube model, the corners of the $m$ loops fuse into fractons if the loops are oriented in the $xz$ or $yz$ planes. Otherwise, they fuse into the vacuum.}
    \label{fig:mfusionlineonani}
\end{figure}

Similarly to the previous model, the fusion of flux excitations gives immobile point excitations at its corners. Interestingly, if the $m$-loop is oriented in the $xy$ plane, then two $m$-loops fuse to the vaccuum. However, if the $m$-loop is oriented in the $xz$ or $yz$ planes, then there will be fracton excitations left at the corners after the fusion. This is shown in Figure \ref{fig:mfusionlineonani}.

The local Wilson operators are products of stabilizers of the Hamiltonian, and correspond to closed trajectories of the excitations in the Hamiltonian as summarized in Table \ref{tab:exc_3foliatedlineon}.

In addition to the usual $-1$ phase between $e^2$ and $m$ simulating the toric code, and between $e$ and $m^2$ simulating X-Cube, we also have a braiding phase of $i$ when an $m$-loop moves around the lineons. By realizing that a product of $\bs A_{v}$ is a closed configuration of the $m$-loop, we see that the $e_x$, $e_y$ and $e_z$ obtains a phase of $i$, $-i$, and $-1$ when an $m$-loop is braided around each particle respectively.

\section{Hybrid Haah's Code}\label{Z4Z2Haah}

\begin{figure*}[t!]
    \centering
   \begin{align*}
   \begin{aligned}[c]
        \bs A_v &= \raisebox{-.5\height}{\includegraphics[scale=0.5]{Av.pdf}} & \bs B_{\nablapic} &= \raisebox{-.5\height}{\includegraphics[scale=0.5]{Bnabla.pdf}} &
        \bs A_v^{HC} &= \raisebox{-.5\height}{\includegraphics[scale=0.5]{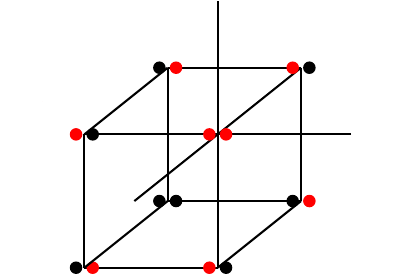}} &
         \bs B_{v}&= \raisebox{-.5\height}{\includegraphics[scale=0.5]{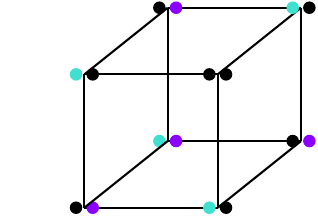}} & 
         \end{aligned}\\
         \begin{aligned}[c]
         \bs \xi_e &=\begin{cases}
         \raisebox{-.5\height}{\includegraphics[scale=0.5]{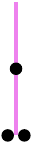}} \vspace{5pt}\\
         \raisebox{-.5\height}{\includegraphics[scale=0.5]{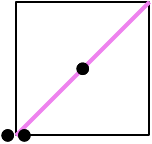}} 
         \end{cases}
         &&\equiv
         \begin{cases}
         \raisebox{-.5\height}{\includegraphics[scale=0.5]{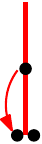}}\vspace{5pt}\\
         \raisebox{-.5\height}{\includegraphics[scale=0.5]{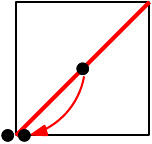}}
         \end{cases} &\bs \xi_e^2 &=
         \begin{cases}
         \raisebox{-.5\height}{\includegraphics[scale=0.5]{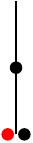}}\vspace{5pt}\\
         \raisebox{-.5\height}{\includegraphics[scale=0.5]{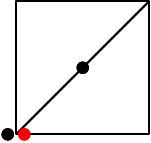}}
        \end{cases}
        \end{aligned}
        \begin{aligned}[c]
         \bs \zeta^{(1)}_v &= \raisebox{-.5\height}{\includegraphics[scale=0.5]{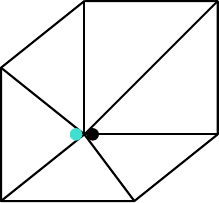}} &&\equiv \raisebox{-.5\height}{\includegraphics[scale=0.5]{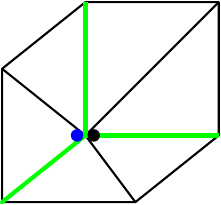}} & \bs \zeta_v^2 &= \raisebox{-.5\height}{\includegraphics[scale=0.5]{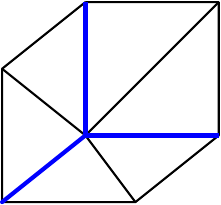}}\\
        \bs \zeta^{(2)}_v &= \raisebox{-.5\height}{\includegraphics[scale=0.5]{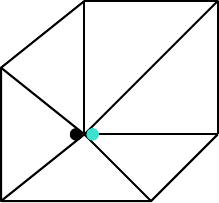}} &&\equiv \raisebox{-.5\height}{\includegraphics[scale=0.5]{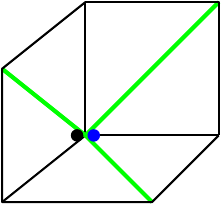}} & \bs \zeta_v^2 &= \raisebox{-.5\height}{\includegraphics[scale=0.5]{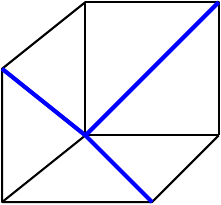}}
        \end{aligned}
        \end{align*}
    \caption{{\bf Lattice Model for the Hybrid Haah's code:} Visualization of the operators in the hybrid Haah's code.  The color coding used is red=$X$, blue=$Z$, green = $S$, magenta = $\bs \xi$, orange = $\bs \xi^\dagger$, cyan = $\bs \zeta$, purple = $\bs \zeta^\dagger$, {\color{red} $\rightarrow$}= \textsc{CNOT}.}
    \label{fig:Z4Z2Haah}
\end{figure*}
In this section, we construct a type-II hybrid fracton model. Here we choose the subsystem symmetry to be the fractal symmetry corresponding to Haah's code \cite{Haah2011}. Therefore, we will call this model the hybrid Haah's code. The symmetric, short-range-entangled state we start with has $\ZZ_4$ d.o.f. on vertices of a cubic lattice. The SRE state is again the ground-state of the Hamiltonian 
\begin{align}
    H =& -\sum_v [ \cX_v+ \cX_v^2 + \cX_v^3].
\end{align}
We enforce a global $\ZZ_4$ symmetry generated by $\prod_v \cX_v$ and a $\ZZ_2$ fractal symmetry $\prod_{v \in \text{fractal}} \cX^2$, the latter of which is precisely the symmetry that is gauged to obtain Haah's code \cite{Haah2011}. The fractal symmetry replaces the planar symmetry previously considered when constructing the hybrid X-Cube orders.

Excitations above the paramagnet ground state are obtained by applying the following operators 
\begin{align}
    \Delta_e &= \cZ_i^2 \cZ_f^2,\\
    \Delta_{v}^{(1)} &=\cZ_{v}\cZ_{v+\hat x}\cZ_{v+\hat y}\cZ_{v+\hat z},\\
    \Delta_{v}^{(2)} &=\cZ_{v}\cZ_{v+\hat x+\hat y}\cZ_{v+\hat y+\hat z}\cZ_{v+\hat x+\hat z}.
\end{align}
where the unit vectors $\hat x$, $\hat y$ and $\hat z$ denotes translation by a unit cell in the $x$, $y$, and $z$ directions respectively.

Next, we follow through the process of gauging the symmetry. We first gauge the $\ZZ_2$ fractal symmetry to obtain Haah's code where the gauge charge is fractionalized by the remaining global $\ZZ_2$ symmetry. Then, we gauge the global $\ZZ_2$ symmetry to obtain the hybrid model.

The hybrid model lives on the same lattice as in Figure \ref{fig:ordering}, and we adopt the same convention of ordering of vertices on each edge. The Hilbert space of this model, however, is different. We place a qubit on each edge and two qubits on each vertex of this lattice. Furthermore, it is helpful to define an index $\alpha=1,2$. On edges, $e^{(1)}$ and $e^{(2)}$ denotes upright or diagonal edges of this lattice, respectively, while on vertices, $v^{(1)}$ and $v^{(2)}$ denote the first and second qubit of that vertex, respectively. 

The algebra of this model is generated by $\bs \zeta^{(\alpha)}_v$, $\bs \xi^{(\alpha)}_e$, $Z_e$, $X^{(\alpha)}_v$. $\bs \zeta_p$ and $\bs \xi_e$ are modified operators given by
\begin{align}
    \bs \xi^{(\alpha)}_e &= X_e \textsc{CNOT}_{e,i_e^{(\alpha)}},
\label{equ:xi_eHaah}\\
\bs \zeta^{(\alpha)}_v&=Z^{(\alpha)}_{v} \prod_{e^{(\alpha)} \leftarrow v} S_e.\label{equ:zeta_vHaah}
\end{align}
Here, $e\leftarrow v$ denotes edges incoming to $v$ as defined in Fig \ref{fig:ordering}.

The operators satisfy the following algebra
\begin{align}
    \bs \zeta^{(\alpha)}_v X^{(\alpha')}_{v'} &= (-1)^{\delta_{v,v'}\delta_{\alpha,\alpha'}} X^{(\alpha')}_{v'}\bs \zeta^{(\alpha)}_v,\\
    \bs Z_e \bs \xi^{(\alpha)}_{e'} &=  (-1)^{\delta_{e,e'}}  \bs \xi^{(\alpha)}_{e'} Z_e,\\
    [\bs \xi^{(\alpha)}_e, \bs \xi^{(\alpha')}_{e'}] &= [\bs \zeta^{(\alpha)}_p, \bs \zeta^{(\alpha')}_{p'}]=[\bs \xi^{(\alpha)}_e, X_p] = [\bs \zeta_p^{(\alpha)}, Z_e]=0.
\end{align}
which makes them look like $\ZZ_2$ Pauli operators when defined solely within the vertex or edge subspace, except that we also have the following commutation relations between terms between vertices and edges
\begin{align}
     \bs \zeta^{(\alpha)}_v \bs \xi^{(\alpha')}_{e} =\begin{cases}
    i \bs \xi^{(\alpha')}_{e}  \bs \zeta^{(\alpha)}_v;  & \text{if } e \leftarrow v \text{\ and \ } \alpha=\alpha', \\
    \ \ \bs \xi^{(\alpha')}_{e}  \bs \zeta^{(\alpha)}_v; & \text{otherwise}.
    \end{cases} 
    \label{equ:xiezetapalgHaah}
\end{align}

The Hamiltonian of the hybrid Haah's code is then expressly
\begin{align}
H_{\text{Hybrid}} =& H_{TC}' + H_{HC}', \nonumber\\
H_{TC}' =&-\sum_v \frac{\bs A_v+\bs A_v^\dagger}{2}  -  \sum_{\nablapic} \frac{1+\bs B_{\nablapic}}{2}, \nonumber \\
     H_{HC}' =& -\sum_v\frac{1+\bs A_v^{HC}}{2} - \sum_{v} \frac{\bs B_{v} + \bs B_{v}^\dagger}{2},
    \label{equ:Z4Z2Haah}
\end{align}
where 
\begin{align}
\bs A_v &=\prod_{e \rightarrow v}\bs \xi_e^\dagger \prod_{e \leftarrow v}\bs \xi_e,\label{equ:AvHaah},\\ \bs B_{{\nablapic}} &=\prod_{e \in {\nablapic}} Z_e,
\label{equ:BnablaHaah}\\
\bs A_v^{HC} &=X^{(1)}_{v}X^{(1)}_{v-\hat x}X^{(1)}_{v-\hat y}X^{(1)}_{v-\hat z}X^{(2)}_{v}X^{(2)}_{v-\hat x-\hat y}X^{(2)}_{v-\hat y -\hat z}X^{(2)}_{v-\hat z -\hat x}, \label{equ:Av2Haah}\\ 
\bs B_{v}&= \bs\zeta^{(1)}_{v}\bs\zeta^{(1)}_{v+\hat x+\hat y}\bs\zeta^{(1)}_{v+\hat y +\hat z}\bs\zeta^{(1)}_{v+\hat z +\hat x}{\bs\zeta^{(2)}_{v}}^\dagger{\bs\zeta^{(2)} _{v+\hat x}}^\dagger {\bs\zeta^{(2)}_{v+\hat y}}^\dagger{\bs\zeta^{(2)}_{v+\hat z}}^\dagger.
\label{equ:BcHaah}
\end{align}
The operators are visualized in Figure \ref{fig:Z4Z2Haah}.

Similarly to previous examples, this model is a hybrid model of a $\ZZ_2$ toric code and a $\ZZ_2$ Haah's code. If we had replaced $\bs \xi_e^{(\alpha)}$ and $\bs \zeta_v^{(\alpha)}$ in the definitions of $\bs A_v$ and $\bs B_{v}$ with Pauli matrices $X_e$ and $Z_v^{(\alpha)}$, respectively, then the model would be a tensor product of the 3d toric code defined on the edges and Haah's code defined on the vertices. However, by replacing $X_e\rightarrow \bs \xi_e^{(\alpha)}, Z_v^{(\alpha)}\rightarrow \bs \zeta_v^{(\alpha)}$, the degrees of freedom on the edges and vertices are now coupled. The vertex term $\bs A_v$ now squares to $\bs A_v^{HC}$ in Haah's code. At the same time, the $\bs B_v$ term of Haah's code now squares to a product of twelve $\bs B_{\nablapic}$ terms of the toric code.

\subsection{Summary of excitations, fusion, and braiding}
The fracton $e$ is created via the operators $\bs \zeta_v^{(\alpha)}$, which commutes with all $\bs B_{v}$ and $\bs B_{\nablapic}$, but violates four $\bs A_v$ projectors. in particular they are charged $i$ under the operator $\bs A_v$. As in Haah's code, a general charge configuration can be created at the corners of a Sierpinski pyramid by applying $\bs \zeta_v^{(\alpha)}$ in a fractal pattern
\begin{align}
    \text{fracton}~e:~~~ \prod_{v \in \text{fractal}} \bs \zeta_v^{(\alpha)} .
\end{align}
The mobile charge can be created at the end points of a string of $Z_e$, which is charged $-1$ under $\bs A_v$. Furthermore, they can also be created by applying $\left(\bs \zeta_v^{(\alpha)}\right)^2$ in a fractal pattern
\begin{align}
    \text{mobile}~e^2:~~~\prod_{v \in \text{fractal}} \left(\bs \zeta_v^{(\alpha)}\right)^2,  \left(\bs \zeta_v^{(\alpha)}\right)^2 = \prod_{e^{(\alpha)} \leftarrow v} Z_e.
\end{align}
Therefore, $e^2$ is a mobile charge.

Next, we define operators that create the flux excitations. First, to create a loop excitation, we apply $\bs \xi_e$ on edges with intersect a surface $\mathcal S'$ in the dual lattice.
\begin{align}
    \text{mobile loop}~m:~~~ \prod_{e\perp \mathcal S'} \bs \xi_p.
\end{align}
The boundary of this surface is charged $-1$ under $\bs B_{\nablapic}$. Furthermore, we find that depending on the shape of $\mathcal S$, the loop is also sporadically charged $i,-1,$ or $-i$ under various $\bs B_v$ operators at the boundary. 

Finally, applying $X_v^{(\alpha)}$ for $\alpha=1,2$ violates four $\bs B_v$ terms, and creates flux fracton excitations $m^2$.
\begin{align}
    \text{fracton}~m^2:~~~ \prod_{v \in \text{fractal}} X_v^{(\alpha)}.
\end{align}

\begin{figure}
    \centering
    \includegraphics[scale=0.7]{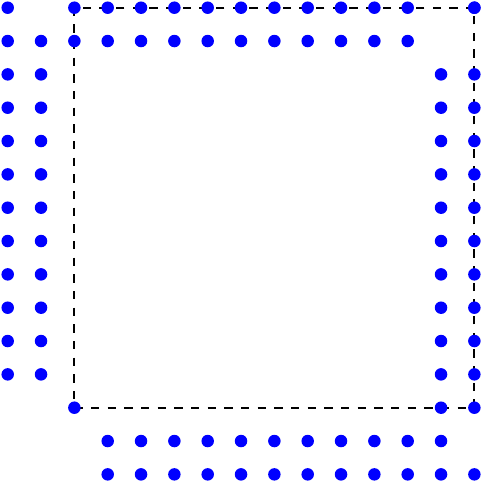}
    \caption{{\bf Loop fusion in the hybrid Haah's code:} Fusing two identical $m$ loops positioned at the dashed square in the $yz$ plane results in fractons ($m^2$) at the positions shown in blue.}
    \label{fig:mfusionHaah}
\end{figure}

The fusion follows from the definition of the operators above. A fusion of two charge fractons $e$ results in a mobile particle $e^2$. Furthermore, a fusion of two $m$-loops results in a number of $m^2$ flux fractons sporadically placed around the loop depending on its shape. For example, the fusion of two $m$-loops on a square in the $yz$ plane is shown in  Fig. \ref{fig:mfusionHaah}.

Braiding is very similar to the hybrid X-Cube models. In addition to the usual $e^2$ and $m$ particle-loop braiding as in the toric code, an $m$-loop can also braid around the fracton $e$ to give a phase of $i$. Unfortunately, we are unaware of well-defined braiding processes between $e$ fracton and $m^2$ fracton in Haah's code. Such types of braiding, if they exist, could give further braiding processes in this model as well.

\section{Hybrid Phases as Parent Orders for Toric Code and Fracton Orders}\label{sec:parent}
In this section, we argue that the hybrid orders introduced in this paper are natural parent states for both liquid and fracton topological orders. We tailor the discussion in this section towards the fractonic hybrid X-Cube model in Sec. \ref{3foliatedZ4Z2}, and demonstrate that driving a phase transition that condenses an appropriate set of excitations in this hybrid order can yield either the $\ZZ_2$ toric code or the $\ZZ_2$ X-Cube model. While we focus on this particular example here, similar arguments can be drawn for all the other models presented in this work. All of the models described in this paper are proximate to the  $\ZZ_2$ toric code or a $\ZZ_2$ fracton order through a similar phase transition.

We may study phases proximate to the fractonic hybrid X-Cube model by adding perturbations to the Hamiltonian $H_\text{Hybrid}$ in Eq. \eqref{equ:Z4Z2gaugetheorysimplified}. We first add longitudinal and transverse fields which act as hopping terms for the mobile charge $e^2$ and the lineon $m^2$ in the hybrid order. 
\begin{align}
    H=H_\text{Hybrid}-t_{e^2}\sum_{e}Z_e - t_{m^2}\sum_p X_p.
    \label{equ:phasediagram}
\end{align}
First, let $t_{e^2}=0$ and consider the limit of large $t_{m^2}$. Since the operator $X_p$ hops a  lineon ($m^{2}$), the lineons ``condense" in the limit of large $t_{m^{2}}$.  As a consequence, the fractons $e$, $\bar{e}$, which braid non-trivially with the lineons are confined\footnote{The energy cost to separate a set of fractons now grows linearly in their separation. This is in contrast to the constant (logarithmic) energy barrier to separate these excitations in a Type I (II) hybrid order.}. The only remaining topological excitations are $e^2$ created by $Z_e$, which is charged $-1$ under $\bs A_v$, and the $m$-loop, created by (\ref{eq:mloop}). Therefore, the $m^2$-lineon condensate is in the same topological phase as the $3d$ toric code.

Indeed, we can derive the effective Hamiltonian in this limit by imposing the constraint that $X_p=1$ on $H_{\text{Hybrid}}$. The operator $\bs B_{c,r}$ does not commute with this constraint and, as a result, does not contribute to the effective Hamiltonian at leading order in perturbation theory in $1/t_{m^{2}}$. The other terms in the Hamiltonian reduce to
\begin{align}\label{eq:replacement_tm2}
    \bs A_v &\rightarrow \prod_{e \supset v} X_e,\\
    \bs B_{\nablapic} &\rightarrow \bs B_{\nablapic} = \prod_{e \in \nablapic} Z_e,\\
    \bs A_{v}^{XC} &\rightarrow 1.
\end{align}
The simplification of the first term follows from the fact that
\begin{align}
\textsc{CNOT}_{e,p} = X_p^\frac{1-Z_e}{2} \rightarrow 1^\frac{1-Z_e}{2}= 1.
\end{align}
The effective Hamiltonian in this subspace, obtained by the replacements in Eq. \eqref{eq:replacement_tm2}, exactly yields the 3d toric code.

\begin{figure}
    \centering
    \includegraphics[scale=0.57]{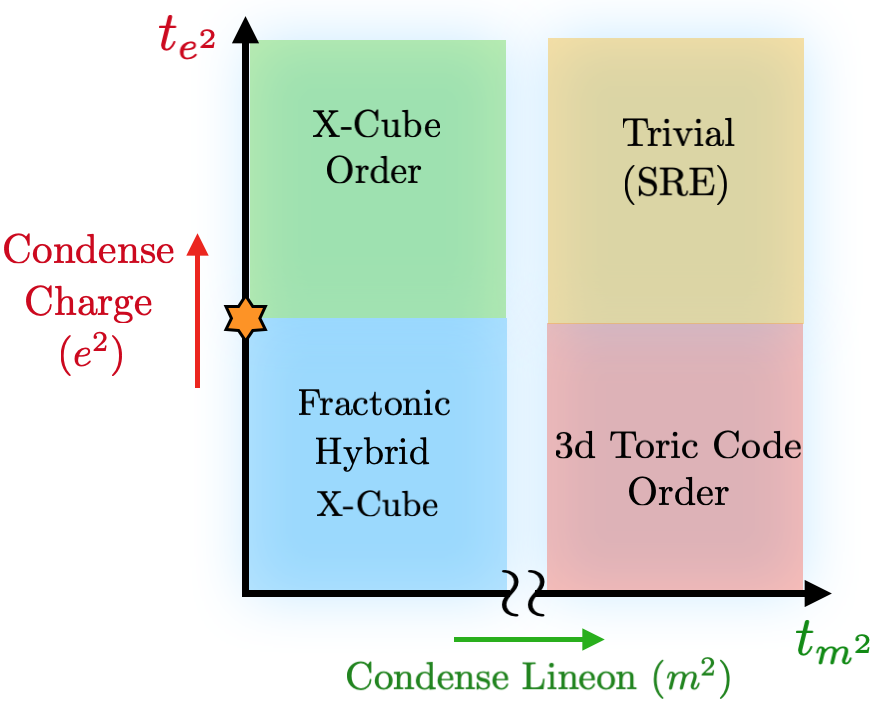}
    \caption{{\bf Schematic Phase Diagram:} For the fractonic hybrid X-Cube model, condensing the mobile charge ($e^2$) or the lineon ($m^{2}$) drives a phase transition into the X-Cube fracton order or a 3d toric code topological order, respectively, as obtained by studying the Hamiltonian \eqref{equ:phasediagram}.  The phase transition between the hybrid order and the X-Cube order in this phase diagram is \emph{continuous} and in the same universality class as the Higgs transition in a three-dimensional $\ZZ_2$ gauge theory, near the line $t_{m^{2}} = 0$, as described in the text. The nature of the phase transition between the hybrid and toric code orders is not known. We note that the geometry of the phase boundaries shown here is not meant to be exact, and we are unaware if other intermediate phases can arise in the ``interior" of the phase diagram. A similar phase diagram is obtained for the hybrid Haah's code in Appendix \ref{app:ParentHaah}.}
    \label{fig:phasediagram}
\end{figure}

Next, we consider the limit that  $t_{m^{2}} = 0$ and $t_{e^2}\rightarrow +\infty$. In this limit, the mobile charge $e^2$ is condensed, as the operator $Z_e$ hops the mobile $e^2$ particle. As a result, the loop excitation $m$, which braids non-trivially with $e^2$ is no longer a topological excitation, and we are left with the fracton $e$ and the lineon $m^2$. They are exactly the excitations in the X-Cube fracton order. More explicitly, we may again obtain the effective Hamiltonian by projecting into the subspace in which $Z_e=1$. The operator $\bs A_v$ brings us out of this constrained subspace and does not contribute to the effective Hamiltonian at leading order. The remaining operators reduce to
\begin{align}
    \bs B_{\nablapic} &\rightarrow 1,\\
    \bs A_{v}^{XC} &\rightarrow \bs A_{v}^{XC} = \prod_{p \supset v} X_p,\\
    \bs B_{c,r} & \rightarrow \prod_{p \in c_r,c_r'} Z_p.
\end{align}
The simplification of the last line follows from
\begin{align}
    S_e = i^\frac{1-Z_e}{2} \rightarrow i^\frac{1-1}{2}=1.
\end{align}
Therefore, the remaining stabilizers are exactly those of the X-Cube model, and the ground-state exhibits the X-Cube fracton order.

Finally, we can consider $t_{e^2}, t_{m^2}\rightarrow \infty$. In this case both $e^2$ and $m^2$ are condensed, since $e$ and $m$ have non-trivial braiding with the above set of excitations, there are no topological excitations left. The phase is a trivial confined phase. A schematic phase diagram is shown in Fig. \ref{fig:phasediagram}.

We may derive some properties of the phase transition between the fractonic hybrid X-Cube and X-Cube orders by observing that the operator $Z_e$ hops the mobile $e^2$ excitation and commutes with all of the terms in the fractonic hybrid X-Cube Hamiltonian, except for $\bs A_v$. Furthermore, the $X_p$ term, which hops a lineon, commutes with all terms in the fractonic hybrid X-Cube Hamiltonian, except for $\bs B_{c,r}$.  As a result, we may study the Hamiltonian in Eq. (\ref{equ:phasediagram}) within a constrained Hilbert space, within which there are no $m$ flux loop excitations, or $e$ fracton excitations, as enforced by $\bs B_{\nablapic} = 1$ and $\bs A_{v}^{XC} = \bs A_{v}^{2} = 1$. Projecting the Hamiltonian into this constrained subspace yields 
\begin{align}\label{eq:projected_Hamiltonian}
    PHP = &-\sum_{v}\bs A_{v} - t_{e^{2}}\sum_{e}Z_{e}\nonumber\\
    &-\sum_{c,r}\bs B_{c,r} - t_{m^{2}}\sum_{p}X_{p},
\end{align}
with $P$ denoting the projection. Here, we have also used the fact that $P \bs A_{v}^{\dagger} P = \bs A_{v}$ and $P\bs B_{c,r} P = \bs B_{c,r}$. 

When $t_{m^{2}} = 0$, the lineons are non-dynamical, and we may further set $\bs B_{c,r} = 1$. The algebra satisfied by $\bs A_{v}$ and $Z_{e}$ is precisely the algebra between the ``star" operator in a 3d toric code, which measures the $\ZZ_2$ charge, and a transverse field, which has the effect of hopping the charge excitation.  As a result, the phase transition between the fractonic hybrid X-Cube order and the X-Cube order is precisely related to the \emph{Higgs transition} in a three-dimensional $\ZZ_2$ gauge theory, which is driven by the condensation of the $\ZZ_2$ charge.  This transition is known to be direct and continuous, and dual to an Ising symmetry-breaking phase transition in (3+1)-dimensions \cite{Wegner1971}, if the dynamical $\ZZ_2$ flux excitations are forbidden.  

The generic nature of the transition between the fractonic hybrid X-Cube and X-Cube orders, when other excitations (fractons or lineons) are allowed is not known.  We may show, however, that the phase transition remains continuous when $t_{m^{2}}$ is turned on. In this case, lineons may be created, though they are highly ``massive" excitations when $t_{m^{2}}$ is small, and we may integrate out the lineon  excitations to obtain an effective description of the critical point. To leading order in perturbation theory, the effective Hamiltonian describing the system when $t_{m^{2}}\ll 1$ may be obtained from Eq. (\ref{eq:projected_Hamiltonian}) as
\begin{align}
    H_{\mathrm{eff}} = &-\sum_{v}\bs A_{v} - t_{e^{2}}\sum_{e}Z_{e}\nonumber\\
    &-\sum_{c,r}\bs B_{c,r} - K\sum_{v}\bs A_{v}^{2} + \cdots
\end{align}
 to leading order in perturbation theory in $t_{m^{2}}$, where $K\sim O(t_{m^{2}}^{12})$.  Since $\bs A_{v}^{2}$ and $\bs B_{c,r}$ commute with the effective Hamiltonian, the critical point separating the fractonic hybrid X-Cube order and the X-Cube model is unchanged at this order in perturbation theory, and the transition remains continuous, and admits a description as a Higgs transition in a three-dimensional $\ZZ_2$ gauge theory.

In Appendix \ref{app:ParentHaah}, we demonstrate through a similar derivation that the Hybrid Haah's code is a parent state for both the toric code and Haah's code. Namely, a condensation of the fracton $m^2$ drives the system into the toric code phase, and a condensation of the mobile charge $e^2$ drives the system into the Haah's code phase.

\section{Discussion}
In this work, we introduced exactly solvable models of hybrid fracton phases, which consists of excitations with both mobile and immobile excitations. As phases of matter, they are distinct from a tensor product of a fracton phase with a liquid topological ordered phase because of its unusual fusion and braiding properties. Our work raises a challenge to classify gapped quantum phases in terms of liquids and non-liquids.

Although in this paper, we focused on Abelian hybrid models, an exactly solvable model can in fact be constructed for an arbitrary finite gauge group, which is presented in follow-up work \cite{TJV2}.

In the following, we present some interesting directions for further studies of Abelian hybrid fracton models. 

{\bf \emph{Twisted hybrid fracton models}:}  In this paper, our hybrid fracton models are obtained by starting with a product state and gauging the $(G,N)$ symmetry. If we instead start with a Subsystem Symmetry Protected Topological (SSPT) state protected by $G$ subsystem symmetry\cite{Devakuletal2018,ShirleySlagleChen20,DevakulShirleyWang2019}, and break the symmetry explicitly to $(G,N)$ it should be possible to obtain more exotic twisted hybrid models after gauging. One might also be interested in finding phases protected directly by the $(G,N)$ symmetries. This can be broadly searched by studying consistency conditions of the symmetry defects \cite{TantivasadakarnVijay2019}.
    
{\bf \emph{Fermionic $(G,N)$ symmetries}:} The current construction can be straightforwardly generalized to fermionic symmetries.  One can start with a charge-$2n$ superconductor, which is in a atomic insulating phase and impose an extra fermion parity symmetry on subregions. Then, one can gauge the fermionic subsystem symmetry \cite{Tantivasadakarn20,Shirley2020} to obtain a $\ZZ_2$ fracton model enriched by a $\ZZ_n$ global symmetry and study whether the symmetry enrichment is different from its bosonic counterpart. If so, further gauging the $\ZZ_n$ global symmetry will give rise to a different hybrid model.
    
{\bf \emph{Subsymmetry-Enriched Topological (SSET) phases}:}  Our hybrid fracton models can be used to construct SSET phases \cite{StephenGarre-RubioDuaWilliamson2020} by condensing the particle $m^2$ along with charges of an Ising model with the same restricted mobility (or equivalently by gauging the higher-rank symmetry associated to the Wilson operator of $m^2$). In particular, using the models introduced in this paper, we expect the resulting model to be a $\ZZ_2$ toric code enriched by $\ZZ_2$ subsystem symmetries. It would be particularly interesting to investigate the properties a $\ZZ_2$ toric code enriched by the fractal symmetry of Haah's code.

{\bf \emph{Error Correcting Codes}:} 
The exotic mix of immobile and mobile excitations in these models might be useful for quantum error correction. Indeed, the absence of string-like logical operators (topologically non-trivial Wilson loops) in Type II fracton orders is intimately related to their improved ability to act as a self-correcting quantum memory \cite{BravyiHaah2013}.  In the hybrid Haah's code, the presence of a loop excitation as well as a conjugate fracton excitation that precisely resembles the fractons in Haah's code may lead to improved performance as a quantum memory, which deserves further study. 

Recently, it has been shown that the topological orders that are Calderbank-Shor-Steane (CSS) codes\footnote{A CSS code is given by a Hamiltonian composed of operators that are either a product of Pauli-$X$ operators, or Pauli-$Z$ operators.}, such as the 3d toric code, X-cube model, or Haah's code do not survive at finite temperature, diagnosed by either the topological entanglement entropy \cite{castelnovo2008topological,li2019finite} or the entanglement negativity\cite{lu2020detecting}. On the other hand, the fractonic hybrid X-cube model introduced in this paper is not a CSS code. It is therefore interesting to see whether the topological order of this model can survive at finite temperature.

{\bf \emph{Emergent symmetries.}} The excitations of the fractonic hybrid X-Cube model in Sec. \ref{3foliatedZ4Z2} seem to have a cubic symmetry, even though the Hamiltonian does not. It would be interesting to see whether the Hamiltonian can be written in a form that preserves the cubic symmetry as well, or whether the cubic symmetry can only emerge at low energies.

{\bf \emph{Non-liquid orders beyond hybrid models.}} As a third proximate phase to the hybrid models, as discussed in Sec. \ref{sec:parent}, it would be interesting to consider condensing the composite excitation $e^2m^2$ in the hybrid model. Such a phase may potentially realize a non-liquid phase that is not a hybrid between a topological order and a fracton order.

\begin{acknowledgments}
The authors are grateful to Yu-An Chen and Tyler Ellison for very helpful discussions on symmetry fractionalization and to Ashvin Vishwanath and Yahui Zhang for interesting discussions on realizing fracton models from condensations. The authors thank David Aasen, Ho Tat Lam, Shu-Heng Shao, and Cenke Xu for stimulating discussions. NT is supported by NSERC. WJ is supported by the Simons Foundation. This work was also partially supported by the Simons Collaboration on Ultra-Quantum Matter, which is a grant from the Simons Foundation.  
\end{acknowledgments}

\appendix

\section{Abelian 2-subsystem symmetries}\label{app:2subsystemdef}

In this Appendix, we give a proper definition of the $(G,N)$ symmetry which act as the gauge group for the hybrid fracton models. Here, we focus on Abelian $(G,N)$ symmetries and give a proper definition for a general group $G$ in Ref. \cite{TJV2}.

\subsection{Definition}

The hybrid fracton models derived in this paper can be thought of as the result of gauging a paramagnet with a particular mix of global and subsystem symmetries, which we term a \emph{2-subsystem symmetry}. This name is an homage to 2-group symmetries\cite{KapustinThorngren2017,DelcampTiwari18,CordovaDumitrescuIntriligator19,BeniniCordovaHsin19}, which can be thought of as a particular extension of global (0-form) symmetries by 1-form symmetries. Likewise, the relevant 2-subsystem symmetries in this paper can be thought of as a global $\ZZ_2$ symmetry extended by a subsystem symmetry corresponding to the underlying fracton order. The hybrid fracton orders can therefore be understood as a ``2-subsystem gauge theory".

An Abelian 2-subsystem symmetry can be defined given the following data
\begin{enumerate}
    \item An Abelian global symmetry group $G$ 
    \item A normal subgroup $N \triangleleft G$
    \item The type of region on which $N$ acts as a subsystem symmetry (e.g. 1-foliated, 3-foliated, fractal,...)
\end{enumerate}
To realize the 2-subsystem symmetry in a lattice model, we first define $R^g_v$ an onsite representation of $g \in G$ at each vertex $v$. The global $G$ symmetry is defined as $\prod_v R^g_v$. Furthermore, for a group element $n$ in the normal subgroup $N$ the $N$ subsystem symmetry is defined as $\prod_{v \in \text{sub}} R^n_v$, where ``sub" is a subregion specified by the type of subsystem symmetry. 

Let us first emphasize two important points. First, since $N$ is a subgroup of $G$, the $N$ subsystem symmetry is not an independent symmetry from the global symmetry. For example, if the subsystem symmetry is 1-foliated, then a product of all $N$ planar symmetries is a global $N$ symmetry, a subgroup of the global $G$ symmetry.

Second, a 2-subsystem symmetry contains both global symmetries and subsystem symmetries as limiting cases. For example, a pure global $G$ symmetry can be written as $(G,\ZZ_1)$ since it only has a trivial subgroup acting as a subsystem symmetry. Furthermore, a pure $N$ subsystem symmetry can be denoted $(N,N)$. It is worth nothing that gauging the symmetries in these two examples will lead to 3d topological orders and fracton orders, respectively, which are not hybrid models. Therefore, the 2-subsystem symmetries we are interested in, are those for which $N$ is a strict but non-trivial normal subgroup of $G$. For example, the gauge group corresponding to the Hybrid toric code layers in Sec. \ref{1foliatedZ4Z2} is a 1-foliated $(\ZZ_4,\ZZ_2)$ subsystem symmetry.

Mathematically, if we neglect the input geometry of the subsystem symmetry (which specifies the region on which the subsymmetry acts), then $(G,N)$ has a structure of an Abelian group. Starting with Abelian groups $G$ and $N$, the central extension
\begin{equation}
1 \rightarrow N  \xrightarrow[]{} G  \xrightarrow[]{} Q \rightarrow 1
\label{equ:exact}
\end{equation}
is given uniquely by specifying a representative cocycle $\omega$ of $H^2(Q,N)$, where $Q=G/N$ is the quotient group.

Now, let $L$ be the number of independent $N$ subsystem symmetries. The group $(G,N)$ is then formally given by the central extension
\begin{equation}
1 \rightarrow N^L  \xrightarrow[]{} (G,N)  \xrightarrow[]{} Q \rightarrow 1
\label{equ:exact2}
\end{equation}
where $N^L \equiv N_1 \times \cdots \times N_L$ is a product of $L$ identical copies of $N$. The extension can be specified by a representative cocycle $\omega'$ of $H^2(Q,N^L)$ chosen to be $ \omega' = \prod_{l=1}^L \iota_l \circ \omega $ where $\iota_l: N \rightarrow N^L$ is the embedding of $N$ to the $l^\text{th}$ copy of $N^L$.

It is important to note that even though $(G,N)$ is a finite Abelian group, and therefore can be written as a product of cyclic groups, specifying the group extension \eqref{equ:exact2} encodes information about which subgroup of $(G,N)$ acts as subsystem symmetries.

\subsection{Gauging 2-subsystem symmetries}
\begin{figure*}[t!]
    \centering
    \includegraphics[scale=0.5]{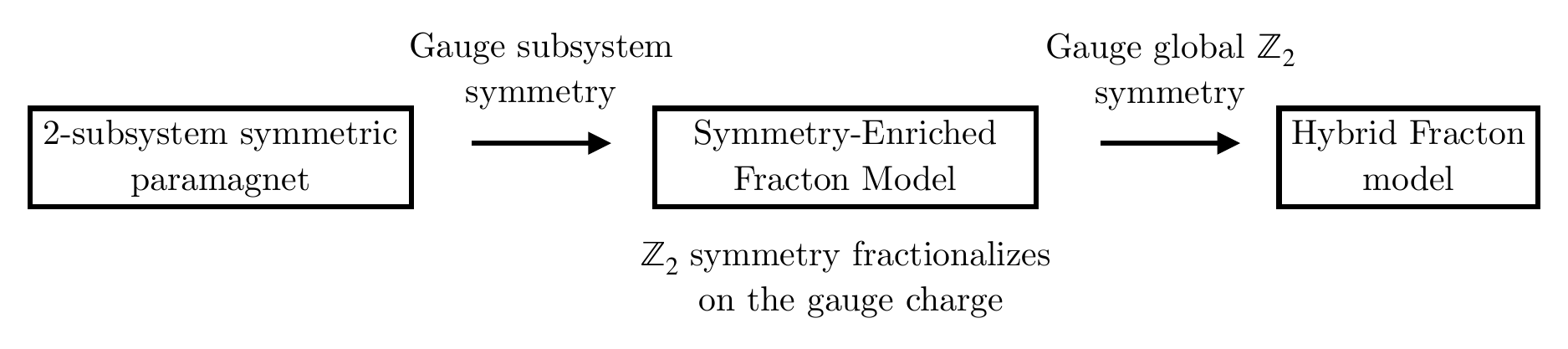}
    \caption{Summary of the gauging process to obtain the Abelian hybrid fracton models in this paper.}
    \label{fig:gauging}
\end{figure*}

The process of gauging a 2-subsystem symmetry to obtain a hybrid fracton model can be broken into two steps, as shown in Figure \ref{fig:gauging}. We start by first gauging the subsystem symmetry $N$, to obtain a fracton model\cite{CobaneraOrtizNussinov2011,VijayHaahFu2016,Williamson2016,KubicaYoshida2018,Pretko2018,ShirleySlagleChen2019,Radicevic2019} (with gauge group $N$). After gauging, the global symmetry $G$ is reduced to the quotient group $Q=G/N$. Nevertheless, because of the interdependence of the global and subsystem symmetries, the remaining global symmetry $Q$ enriches the fracton phase. In particular, for Abelian 2-subsystem symmetries, $Q$ fractionalizes on the gauge charge (which has restricted mobility) of the fracton model. The fractionalization implies that these gauge charges can fuse into charges of $Q$. Now, since gauging $Q$ promotes the charges of $Q$ into gauge charges, the result is a gauge theory where gauge charges with restricted mobility can fuse into fully mobile gauge charges. This is precisely our hybrid fracton model. Furthermore, consistency with braiding implies that the flux loops should also square into various fluxes with restricted mobility. These properties are demonstrated explicitly in our exactly solvable lattice models.

The explicit process of gauging 2-subsystem symmetries is omitted from the main text. Instead, we give a detailed calculation of gauging a 2-subsystem symmetry that gives the (fractonic) hybrid X-Cube model of Sec. \ref{3foliatedZ4Z2} in Appendix \ref{app:Z4Z2derivation}.

\section{Derivation of the fractonic hybrid X-Cube model}\label{app:Z4Z2derivation}

In this appendix, we present details of the derivation of the fractonic hybrid X-Cube model. We first present the form of the corresponding Ising model with $(\ZZ_4,\ZZ_2)$ symmetry. We then gauge the symmetry to obtain the corresponding gauge theory. We note that the final model we obtain given in Eq. \eqref{equ:Z4Z2gaugetheory} is slightly different from Eq. \eqref{equ:Z4Z2gaugetheorysimplified} in the main text, but only differs in the energetics. That is, the stabilizers of the two models are the same. A prescription for gauging a general 2-subsystem symmetry is presented in Ref. \onlinecite{TJV2}. 

\subsection{Ising model}

We begin with a cubic lattice with an additional diagonal edge added to each plaquette as shown in Figure \ref{fig:ordering}. The additional edge, though not essential to the construction, is for convenience. Namely, it makes the minimal coupling to the global symmetry unambiguous.  In addition, it will also make the resulting terms after gauging are non-Pauli stabilizers of the ground state subspace. Lastly it will make the form of the Wilson operators apparent.

On each vertex $v$, we place a $\ZZ_4$ qudit associated to clock and shift operators Eq. \eqref{equ:clockshift} reproduced here
\begin{align}
   \cZ &= 
\begin{pmatrix}
1 & 0 & 0 & 0 \\
0 & i & 0 & 0 \\
0 & 0 & -1 & 0 \\
0 & 0 & 0 & -i \\
\end{pmatrix}, &\cX & = 
\begin{pmatrix}
0 & 0 & 0 & 1 \\
1 & 0 & 0 & 0 \\
0 & 1 & 0 & 0 \\
0 & 0 & 1 & 0 \\
\end{pmatrix}. \label{equ:clockshiftapp}
\end{align}
The Hamiltonian is the following transverse-field Potts model
\begin{align}
    H =& -\sum_v \frac{1+ \cX_v+ \cX_v^2 + \cX_v^3}{4} \nonumber\\
     &- h_\text{E} \sum_{e = (if)} \frac{1+\Delta_e}{2}- h_\text{P} \sum_{p=(ijkl)} \frac{1+\Delta_p + \Delta_p^2 +\Delta_p^3}{4}
\end{align}
where there are two type of Ising terms: one on each edge, and one on each square plaquette
\begin{align}
    \Delta_e &= \cZ_i^2 \cZ_f^2,\\
    \Delta_p &=\cZ_i\cZ_j^\dagger\cZ_k\cZ_l^\dagger.
\end{align}
The Hamiltonian has a global $\ZZ_4$ symmetry generated by $\prod_v \cX_v$, and a planar $\ZZ_2$ symmetry $\prod_{v \in \text{plane}} \cX^2_v$. We note that without $\Delta_e$, the model would instead have a larger symmetry, namely a $\ZZ_4$ symmetry generated by $\prod_{v \in \text{plane}} \cX$ for each plane.

When $h_\text{E}, h_\text{P} \ll 1$, the model is in the paramagnetic phase. For simplicity, we will now analyze the model in the limit where $h_\text{E}= h_\text{P} =0$. In this limit, gapped charged excitations to the ground state are created by applying $\Delta_e$, and $\Delta_p$, the nature of which has been described in the main text.

\subsection{Gauging $\ZZ_2$ planar symmetries}
To gauge the symmetry, it is helpful to first rewrite each vertex as a tensor product of two qubits\footnote{A similar mapping which differs by a Hadamard on the first qubit was pointed out in Ref. \onlinecite{ShirleySlagleChen20}. In Ref. \onlinecite{TJV2} we generalize the mapping presented to any group extension described by a factor system.}. From the expressions in Eq. \eqref{equ:clockshiftapp}, notice that
\begin{align}
\cZ  &= Z \otimes S , &\cX & = (\mathbbm 1 \otimes X) \times \textsc{CNOT}_{2,1},\\
\cZ^2  &= \mathbbm 1 \otimes Z , &\cX^2 & = X \otimes \mathbbm 1.\label{equ:Z4totwoZ2}
\end{align}
where
\begin{align}
    \textsc{CNOT}_{2,1} = \begin{pmatrix}
1 & 0 & 0 & 0 \\
0 & 0 & 0 & 1 \\
0 & 0 & 1 & 0 \\
0 & 1 & 0 & 0 \\
\end{pmatrix}
\end{align}
is the $\textsc{CNOT}$ gate controlled by the second qubit and targetted at the first qubit. Therefore, we see that the planar symmetry $\prod_{v \in \text{plane}} \cX^2_v$ in this basis is just a product of Pauli $X$ operators in the first qubit, and we can perform a generalized Kramers-Wannier duality \cite{VijayHaahFu2016,ShirleySlagleChen2019,Radicevic2019} on the first qubit to gauge the $\ZZ_2$ planar symmetries and obtain the X-Cube model. Under the duality, the first qubit is mapped to qubits on plaquettes of the cubic lattice, while the second qubit remains on each vertex. At the level of operators, it is defined as
\begin{align}
 (Z\otimes I)_i  (Z\otimes I)_j  (Z\otimes I)_k  (Z\otimes I)_l &\rightarrow Z_p,\\
 (X\otimes I)_v &\rightarrow \prod_{p \supset v} X_p.
\end{align}
Note that only operators that commute with the planar symmetries will be mapped to local operators. The result of this mapping on the symmetric operators is
\begin{align}
    \Delta_e &\rightarrow \tilde \zeta_e =Z_i Z_f,\\
    \Delta_p &\rightarrow \tilde \zeta_p =Z_p S_iS_j^\dagger S_k S_l^\dagger,\\
    \cX &\rightarrow  \tilde A_v =X_v \times \prod_{p \supset v} \textsc{CNOT}_{v,p},\\
    \cX^2 &\rightarrow \tilde A_v^2 =\prod_{p \supset v} X_p.
\end{align}
Here, we note the operators with tildes because the remaining global symmetry action is not written in an onsite manner. We will momentarily fix this via a basis transformation and denote operators in the new basis without tildes.

Now, because a product of four $\Delta_p$ operators around a cube is the identity, we also must enforce constraints
\begin{equation}
    \tilde B_{c,r}=\prod_{p \in c_r' } \tilde \zeta_p^\dagger  \prod_{p \in c_r }\tilde \zeta_p = \prod_{p \in c_r, c_r'} Z_p=1
\end{equation}
where $c_r'$ and $c_r$ are the set of plaquettes surrounding a cube defined in Figure \ref{fig:Z4Z2}. This can be enforced energetically via a projector. Therefore, the Hamiltonian after gauging the planar subsystem symmetries is
\begin{align}
    H_{SEF} = &-\sum_v \frac{1+ \tilde A_v+ \tilde A_v^2 + \tilde A_v^3}{4} \nonumber\\
    &-\sum_p \sum_{r=x,y,z} \frac{1+ \tilde B_{c,r}+ \tilde  B_{c,r}^2 + \tilde B_{c,r}^3}{4}.
\end{align}
Looking at the terms defined only on plaquettes, given by $\tilde A_v^2 $ and $\tilde B_{c,r}$ one notices that these are indeed just the terms that appear in the X-Cube model. However, the model is coupled to a $\ZZ_2$ matter field living on vertices in a non-trivial manner.\footnote{To elaborate, the term $\tilde A_v$ is not a minimally coupled term between the X-cube model and a $\ZZ_2$ matter field since $\tilde A_v$ has order four.}

Nevertheless, the model has a global symmetry that enriches the X-Cube fracton order, thus realizing a Symmetry-Enriched Fracton (SEF) model. Since we have gauged the $\ZZ_2$ planar symmetries, the $\ZZ_4$ global symmetry is now reduced to a $\ZZ_2$ global symmetry, generated by $\prod_v \tilde A_v$ (the reason it has order two is because $\prod_v \tilde A_v^2$ is just the identity). However, written in the current basis, the symmetry is not onsite because $\tilde A_v$ acts both on the vertex $v$ as well as nearby plaquettes, and therefore is not a local symmetry action. Indeed, this property allows us to demonstrate that this $\ZZ_2$ symmetry fractionalizes on the fracton excitation, which we will now prove. Consider a fracton excitation at a vertex $v_0$, which is a charge $\tilde A_{v_0}^2=-1$ of the vertex term of the X-Cube model. If we consider a local action of the $\ZZ_2$ symmetry in a region $\mathcal R$, denoted  $\prod_{v\in \mathcal R} \tilde A_v$ such that only this fracton excitation is in the region $\mathcal R$, then the local symmetry action squared is $\prod_{v\in \mathcal R} \tilde A_v^2=-1$. This shows that the fracton is indeed fractionalized under the global $\ZZ_2$ symmetry.

\subsection{Gauging the global $\ZZ_2$ symmetry}
Our next goal is to gauge the remaining $\ZZ_2$ global symmetry. However, in its current form, the global symmetry $\prod_v \tilde A_v$ is not onsite. Therefore, we have to perform a basis transformation to make the symmetry onsite. That is, a basis where the local symmetry action is $X_v$ instead of $A_v$. Our basis transformation is the following unitary
\begin{align}
    U = \prod_p X_p^{\frac{1+Z_i}{2}\left (\frac{1-Z_j}{2} +\frac{1-Z_k}{2} + \frac{1-Z_l}{2} \right )}.
    \label{eqn:unitarytfm}
\end{align}

Using this unitary to conjugate all operators, we find that the operators in the new basis (now written without tildes) are
\begin{align}
A_v =& \prod_{p|v=j,k,l} X_p^{\frac{1-Z_iZ_v}{2}} \times X_v \nonumber \\
&\times \prod_{p|v=i} X_p^{\frac{1-Z_vZ_j}{2} +\frac{1-Z_vZ_k}{2} + \frac{1-Z_vZ_l}{2}},\\
A_v^2 =& \prod_{p \supset v} X_p, \\
B_{c,r}=& \prod_{p \in c_r }\zeta_p, \\
\zeta_e =& Z_iZ_f,\\
\zeta_p =& Z_p (S_i CZ_{ij}S_j) (S_i CZ_{ik}S_k)^\dagger (S_i CZ_{il}S_l).
\end{align}
In particular, the global $\ZZ_2$ symmetry now acts as $\prod_v A_v = \prod_v X_v$, which is onsite, since all the $X_p$ factors cancel in the product.

We can now perform the Kramers-Wannier duality to gauge the global $\ZZ_2$ symmetry. This maps the remaining qubit on each vertex, to qubits on each edge. At the level of operators, the map is given by
\begin{align}
Z_i Z_f &\rightarrow Z_e,\\
S_i CZ_{if} S_f &\rightarrow S_e,\\
X_v &\rightarrow \prod_{e\supset v} X_e.
\end{align}

The gauged operators are now denoted in bold, and are
\begin{align}
\bs A_v =&  \prod_{p|v=j,k,l} X_p^{\frac{1-Z_{iv}}{2}} \times \prod_{e \supset v} X_e \nonumber \\
&\times \prod_{p|v=i} X_p^{\frac{1-Z_{vj}}{2} +\frac{1-Z_{vk}}{2} + \frac{1-Z_{vl}}{2}},\\
\bs A_v^2=&\prod_{p \supset v} X_p, \\
\bs B_{c,r}=& \prod_{p \in c_r }\bs \zeta_p, \\
\bs \zeta_e =& Z_e,\\
\bs \zeta_p =& Z_p S_{ij}S^\dagger_{ik} S_{il}.
\end{align}
We remark that $ X_p^{\frac{1-Z_{e}}{2}} = \textsc{CNOT}_{e,p}$. Therefore, defining $\bs \xi_e$ as in Eq. \eqref{equ:xi_e}, we realize that $\bs A_v$ can be rewritten as in Eq. \eqref{equ:Av}.

Enforcing a fluxless condition using for each triangle $\nablapic$ on the lattice using $\bs B_{\nablapic}$ energetically, we finally obtain the Hamiltonian for the $(\ZZ_4,\ZZ_2)$ gauge theory
\begin{align}
    H =& -\sum_v \frac{1+ \bs A_v+ \bs A_v^2 + \bs A_v^3}{4}  -  \sum_{\nablapic} \frac{1+\bs B_{\nablapic}}{2}  \\
     &- \sum_{c}\sum_{r=x,y,z} \frac{1 + \bs B_{c,r} + \bs B_{c,r}^2 + \bs B_{c,r}^3}{4},
     \label{equ:Z4Z2gaugetheory}
\end{align}
This Hamiltonian has the same ground state and types of excitations as that of the fractonic hybrid X-Cube model in Eq. \eqref{equ:Z4Z2gaugetheorysimplified}, and only differs in the energetics. In particular, we can identify $\bs A_v^2$ with $\bs A_v^{TC}$ in the main text.

\section{Derivation of the lineonic hybrid X-Cube model}\label{sec:pstringlineon}
In this Appendix, we derive the lineonic hybrid X-Cube model. However, instead of deriving this model from gauging a SRE state with particular symmetry like in Appendix \ref{app:Z4Z2derivation}, we will show that Lineonic hybrid X-Cube model can in fact be alternatively be derived via $p$-string condensation\cite{MaLakeChenHermele2017}. We start with the hybrid toric code layers of Sec. \ref{1foliatedZ4Z2} (where the layers are in the $xy$ planes) tensored with stacks of 2d toric codes along the $xz$ and $yz$ planes. The stabilizers of this stacked model are
\begin{align}
    \bs A^{(1)}_{v}&=\raisebox{-0.5\height}{\includegraphics[scale=1]{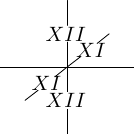}} & 
   \bs B^{(1)}_{p_{xz}}&= \raisebox{-0.5\height}{\includegraphics[scale=1]{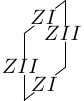}}  \nonumber\\
   \bs A^{(2)}_{v}&=\raisebox{-0.5\height}{\includegraphics[scale=1]{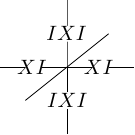}} &
      \bs B^{(2)}_{p_{yz}}& = \raisebox{-0.5\height}{\includegraphics[scale=1]{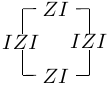}}
\nonumber \\
   \bs A^{(3)}_{v}&= \raisebox{-0.5\height}{\includegraphics[scale=1]{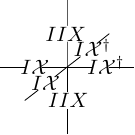}} & 
    \bs B^{(3)}_{p_{xz}}&= \raisebox{-0.5\height}{\includegraphics[scale=1]{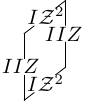}} \nonumber\\
    \bs B^{(3)}_{p_{yz}} &= \raisebox{-0.5\height}{\includegraphics[scale=1]{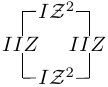}}
    & \bs B^{(3)}_{p_{xy}}& =\raisebox{-0.5\height}{\includegraphics[scale=1]{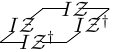}}  \label{equ:stabilizerscondense}
\end{align}
Next we perform a $p$-string condensation, by enforcing the constraint
\begin{align}
(X\cX^2)_e &=1 & \text{(for $x,y$ edges)}\\
(XXI)_e &=1 & \text{(for $z$ edges)}
\end{align}
This condenses the particle-string composed of the $m$ anyons of the toric code stacks in the $xz$ and $yz$ planes, along with the $m^2$ planon of the hybrid toric code layers.

The constraint above reduces the size of the Hilbert space. We can define effective Pauli operators which commutes with the above constraints and satisfy the same algebra as the original ones. For $x$ and $y$ edges, we choose
\begin{align}
I\cX &\equiv \cX, &  I\cX^2 = XI &\equiv \cX^2 & Z\cZ &\equiv \cZ,  
\end{align}
and on $z$ edges, we choose
\begin{align}
XII = IXI &\equiv XI, & IIX &\equiv IX, & ZZI &\equiv ZI, & IIZ &\equiv IZ.
\end{align}

We can now derive the effective stabilizers in this subspace by restricting to only product of the stabilizers in Eq. \eqref{equ:stabilizerscondense}  that commutes with the constraints. The stabilizers restrict to the following stabilizers of the lineonic hybrid X-Cube model under the substitution
\begin{align}
\bs A_v^{(1)} &\rightarrow \bs A_{v,x}^{XC},\\
\bs A_v^{(2)} &\rightarrow \bs A_{v,y}^{XC},\\
\left (\bs A_v^{(3)}\right )^2 &\rightarrow \bs A_{v,z}^{XC},\\
\bs A_v^{(3)} &\rightarrow \bs A_{v,},\\
\bs B^{(3)}_{p_{xz}} &\rightarrow \bs B_{p_{xz}},\\
\bs B^{(3)}_{p_{yz}} &\rightarrow \bs B_{p_{yz}},\\
\left (\bs B^{(3)}_{p_{yz}}\right)^2 &\rightarrow \bs B_{p_{xy}}.
\end{align}
Now, although $\bs B^{(1)}_{p_{xz}}$, $\bs B^{(2)}_{p_{yz}}$, and $\bs B^{(3)}_{p_{xy}}$ do not commute with the constraints, the product of such operators around the six plaquettes surrounding a cube does commute. The restriction of such cube operators is exactly $\bs B_c$ in the lineonic model. Therefore, we have shown that the effective Hamiltonian after the condensation realizes the lineonic hybrid X-Cube model as desired.

We remark that the lineonic $(\ZZ_4^2,\ZZ_2^2)$ hybrid model presented in Appendix \ref{app:Z44Z22lineon} can similarly be derived by starting with a product three copies of the hybrid 1-foliated model presented in Sec. \ref{1foliatedZ4Z2}, oriented in the three directions. The condensate is a product of the three $m$-loops of each model. Interestingly, since $m$-loops of the 1-foliated model square to a product of $m^2$ planons in each layer, the square of the term that condenses the three $m$-loops is exactly the term that induces a $p$-string condensate of the $m^2$ planons. A further study of the relation between flux loops and $p$-strings could shed further light on the relation between higher-form symmetries and their foliated versions\cite{QiRadzihovskyHermele20}.

\section{Ground State Degeneracy}\label{app:GSD}

In this Appendix, we calculate the ground state degeneracy of the hybrid fracton models introduced in the paper. We first perform explicit calculations for the hybrid toric code layers and the fractonic hybrid X-Cube model on a torus, and then prove generally that the ground state degeneracy of the hybrid model is equal to that of the tensor product model.

\subsection{Hybrid Toric Code Layers} \label{app:1foliatedGSD}
The Hamiltonian \eqref{equ:1foliatedHam} has a $\ZZ_2$ qubit and two $\ZZ_4$ qubits per unit cell. Therefore,
\begin{align}
    \log_2 (\dim \mathcal H) = 5L_xL_yL_z
\end{align}
Since the Hamiltonian is a commuting projector, the ground state subspace are eigenstates for which we set all the stabilizers to one
\begin{align}
    \bs A_v = \bs A_v^{2d}= \bs B^{3d}_p = \bs B_{p_{xy}}=1.
\end{align}
Let us count how many constraints this imposes on the Hilbert space. 

First, there is one $\bs A_v$ operator per unit cell. Since $\bs A_v$ has eigenvalues $\pm 1$ and $\pm i$, restricting to $\bs A_v=1$ divides the size of the Hilbert space by four, meaning it produces two constraints per vertex, and so it imposes $2L_xL_yL_z$ constraints in total. Not all such constraints are independent. For each $xy$ plane, we have the identity $\prod_{v \in xy} \bs A_v^2=1$. There are $L_z$ such identities. Furthermore, we also have the constraint $\prod_v \bs A_v=1$. Note that $\prod_v \bs A_v^2=1$ is already accounted for by the previous identities. Therefore we have \begin{align}
    2L_xL_yL_z -L_z -1
\end{align}
constraints from $\bs A_v$. We note that since $\bs A_v^2 = \bs A_{v}^{2d}$, there are no further constraints from setting $\bs A_{v}^{2d}=1$.

Next, there are three $\bs B_{p}^{3d}$ operators per unit cell. The eigenvalues of $\bs B_{p}^{3d}$ are $\pm 1$, so this produces $3L_xL_yL_z$ constraints. We subtract by $L_xL_yL_z-1$ for each cube where a product of six $\bs B_{p}^{3d}$ operators around the cube is the identity; the extra one is because of overcounting all the cubes, and we subtract by three for the product of $ \bs B_{p}^{3d}$ around the three 2-cycles of the torus being identity. The number of independent $\bs B_{p}^{3d}$ constraints is therefore
\begin{align}
    3L_xL_yL_z - (L_xL_yL_z-1)-3 = 2L_xL_yL_z-2.
\end{align}

Finally, there is one $\bs B_{p_{xy}}$ operator per unit cell, which has eigenvalues $\pm 1$ and $\pm i$. However, $\bs B_{p_{xy}}^2 = \bs B_{p_{xy}}^{3d}$, so this only imposes $L_xL_yL_z$ further constraints. For each plane, we have the identity $\prod_{v \in xy} \bs B_{p_{xy}}=1$, so subtracting the redundancies, there are only
\begin{align}
    L_xL_yL_z -L_z
\end{align}
additional constraints

Putting everything together, the ground state degeneracy is given by
\begin{align}
    \log_2 \textsc{GSD} =& 5L_xL_yL_z - ( 2L_xL_yL_z - L_z-1)\nonumber \\
    &-( 2L_xL_yL_z-2)-  (L_xL_yL_z -L_z)\nonumber \\
    &=2L_z+3
\end{align}
which is the same as the GSD of $L_z$ layers of 2d toric codes tensored with a 3d toric code.

Let us construct these logical operators explicitly. First, in each layer, we can tunnel the $e$ planon around the $x$ cycle of the 2d torus using
\begin{align}
    W^e_{x,l} = \prod_{e_x,y=y_0, z=l} \cZ_e
\end{align}
for some fixed $y_0$. There are $L_z$ such operators for each layer $l=1\ldots L_z$. This anticommutes with
\begin{align}
    W^{m^2}_{y,l} = \prod_{e_x, x=x_0, z=l} \cX_e^2
\end{align}
for some fixed $x_0$. This operators that tunnels the $m^2$ planon around the $y$ cycle of the 2d torus in the layer $l$. Each pair spans a Hilbert space of dimension two, therefore so far they fit in a Hilbert space of dimension $2^{L_z}$.

In addition, we can also consider tunneling the $m$ loop around the $yz$ 2-cycle of the 3d torus.
\begin{align}
    W^{m}_{yz} = \prod_{e_x, x=x_0} \cX_e
\end{align}
This commutes up to a phase $i$ with $W^e_l$ for each $l$. Note that
\begin{align}
  \left ( W^{m}_{yz}\right )^2 = \prod_{l} W^{m^2}_{y,l}.
\end{align}
It turns out that we can minimally extend the size of the Hilbert space by a single qubit to accomodate this operator. That is, the algebra of these operators fit in a Hilbert space of dimension $2^{L_z+1}$. Explicitly, we can express the above operators using $L_z +1$ qubits as
\begin{align}
    W^e_{x,l} &= Z_l S_{L_z+1},\\
    W^{m}_{yz} &= X_{L_z+1} \prod_l \textsc{CNOT}_{L_z+1,l},\\
    W^{m^2}_{y,l} &= X_l,
\end{align}
which satisfies the same algebra. Note that there is another set of operators constructed identically by swapping $x$ and $y$, independent of this set of operators.\\
Finally, we can tunnel the $e^2$ mobile particle in the $z$ direction using
\begin{align}
    W^{e^2}_{z} = \prod_{e_z,x=x_0, y=y_0} Z_e
\end{align}
This anticommutes with 
\begin{align}
    W^{m}_{xy} = \prod_{e_z,z=z_0} X_e
\end{align}
which tunnels $m$ around the 2-cycle in the $xy$ plane. This pair generates a Hilbert space of dimension two independent of the two aforementioned sets. Putting everything together, the Hilbert space dimension of the logical subspace is
\begin{align}
    \log_2 (\dim \mathcal H_\text{logical}) = 2(L_z+1)+1= 2L_z+3
\end{align}
in agreement with our ground state degeneracy.

\subsection{Fractonic Hybrid X-Cube}

The Hamiltonian has nine qubits per unit cell, and therefore lives in a Hilbert space with $\log_2 (\dim \mathcal H) = 9L_xL_yL_z$. Since the Hamiltonian is a commuting projector, the ground state subspace are eigenstates for which we set all the stabilizers to one
\begin{align}
    \bs A_v = \bs A_v^{TC}= \bs B_{\nablapic} = \bs B_{c,r}=1.
\end{align}

First, there is one $\bs A_v$ per unit cell, but since it has eigenvalues $\pm 1$ and $\pm i$, restricting to $\bs A_v=1$ divides the size of the Hilbert space by four, meaning it produces two constraints per vertex, and so it imposes $2L_xL_yL_z$ constraints in total. Not all such constraints are independent. Similarly to X-Cube, we subtract by $L_x+L_y+L_z-2$ because $\prod_\text{plane} \bs A_v^2=1$ for each $xy$, $yz$, and $xz$ plane; the overcount of two is because the product over all parallel planes is $\prod_v \bs A_v^2$. Then, analogously to the toric code, we subtract by one because $\prod_v \bs A_v =1$ ($\prod_v \bs A_v^2=1$ already being accounted just earlier). To conclude there are
\begin{align}
    2L_xL_yL_z - (L_x+L_y+L_z-1)
\end{align}
independent $\bs A_v$ constraints. Since $\bs A_v^2 = \bs A_{v}^{TC}$, there are no further constraints from setting $\bs A_{v}^{TC}=1$

Next, there are six $\bs B_{\nablapic}$ operators per unit cell. The eigenvalues of $\bs B_{\nablapic}$ are $\pm 1$, so this produces $6L_xL_yL_z$ constraints. We subtract by $L_xL_yL_z-1$ for each cube where a product of twelve $\bs B_{\nablapic}$'s around the cube is the identity; the extra one is because of overcounting all the cubes, and we subtract by three for the product of $\bs B_{\nablapic}$ around the three 2-cycles of the torus being identity. The number of independent $\bs B_{\nablapic}$ constraints is therefore
\begin{align}
    6L_xL_yL_z - (L_xL_yL_z-1)-3.
\end{align}

Finally there are two independent $\bs B_{c,r}$ operators per unit cell, each of which has eigenvalues $\pm 1$ and $\pm i$. Thus, there are $4L_xL_yL_z$ constraints. We need to subtract by $L_x+L_y+L_z-1$ because $\prod_\text{plane} \bs B_{c,r}=1$ for each $r$ perpendicular to the plane, the extra one for overcounting the product over all possible planes being identity. Furthermore, each $\bs B_c$ squares to a product of four $\bs B_{\nablapic}$ operators, meaning we have to further subtract by $2L_xL_yL_z$. Therefore, the independent constraints of $\bs B_{c,r}$ up to $\bs B_{\nablapic}$ operators is
\begin{align}
    2L_xL_yL_z -(L_x+L_y+L_z-1)
\end{align}

Putting everything together, the ground state degeneracy is given by
\begin{align}
    \log_2 \textsc{GSD} =& 9L_xL_yL_z - [ 2L_xL_yL_z - (L_x+L_y+L_z-1)]\nonumber \\
    &- [ 6L_xL_yL_z - (L_xL_yL_z-1)-3] \nonumber\\
    &- [ 2L_xL_yL_z -(L_x+L_y+L_z-1)]\nonumber \\
    &=2(L_x+L_y+L_z)
\end{align}
which is consistent with the number of logical operators counted in the main text.

\subsection{General calculation}\label{app:generalGSD}
Let us now show in general that a hybrid model between a $\ZZ_2$ toric code and a $\ZZ_2$ fracton model will have the same ground state degeneracy on any manifold. To warm up, let us calculate the ground state degeneracy in the stacked model. We have stabilizers
\begin{align}
    \left (\bs A^{TC}\right)^2= \left (\bs B^{TC}\right)^2= \left (\bs A^\text{frac}\right)^2= \left (\bs B^\text{frac}\right)^2=1.
\end{align}
We omit the cell which each stabilizers is defined on for simplicity. The dimension of the Hilbert space is given by
\begin{align}
    \log_2 \text{dim } \mathcal H_\text{stack} = \log_2 \text{dim } \mathcal H_\text{TC} + \log_2 \text{dim } \mathcal H_\text{frac}.
\end{align}
The stabilizers of the toric code $\bs A^{TC}$ and $\bs B^{TC}$ imposes $\log_2 \text{dim } \mathcal H_\text{TC}-3$ independent constraints while those of the fracton model $\bs A^\text{frac}$ and $\bs B^\text{frac}$ will impose $\log_2 \text{dim } \mathcal H_\text{frac} -\log_2 \textsc{GSD}_\text{frac}$ constraints. Therefore the ground state degeneracy of the stacked model is
\begin{align}
   \log_2 \textsc{GSD}_\text{stack} &= \log_2 \textsc{GSD}_\text{frac} +3.
\end{align}
We now argue that the ground state degeneracy in the hybrid model must be the same. The dimension of the Hilbert space in the hybrid model is equal to that of the stacked model, while the stabilizers satisfy
\begin{align}
    \left (\bs A^{TC}\right)^2 & = \bs A^\text{frac},\nonumber\\
    \left (\bs B^\text{frac}\right)^2 &= \prod \bs B^{TC}, \nonumber \\
    \left (\bs A^\text{frac}\right)^2 =\left (\bs B^{TC}\right)^2 &=1
\end{align}
where $\prod$ is an appropriate product of toric code plaquette terms, depending on the hybrid model. Because of this, the number of constraints that $\bs A^\text{frac}$ and $\bs B^{TC}$ impose are unchanged, while the \textit{additional} constraints that $\bs A^{TC}$ and $\bs B^\text{frac}$ impose (up to $\bs A^\text{frac}$ and $\bs B^{TC}$ terms) are also the same as before. This implies that the ground state degeneracy of the hybrid model is the same as that of stacked model.

It is important to note that the above result does not imply that the logical operators of the hybrid model satisfy the same algebra to those in the stacked model. Though this is indeed the case in the hybrid X-Cube model, we have demonstrated that the algebra is different in the case of the hybrid toric code layers; some logical operators in the hybrid model have order four, while all logical operators have order two in the stacked model.

\section{A Hybrid of X-Cube and Two Toric Codes}\label{app:Z44Z22lineon}

\begin{figure*}[t]
    \centering
    \begin{align*}
\bs A_{v,1} &= \raisebox{-.5\height}{\includegraphics[scale=0.7]{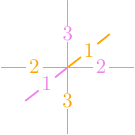}} & \bs A_{v,1}^2 &= \raisebox{-.5\height}{\includegraphics[scale=0.7]{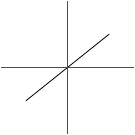}} & \bs B_{p,1}&= \raisebox{-.5\height}{\includegraphics[scale=0.7]{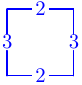}}, &&\raisebox{-.5\height}{\includegraphics[scale=0.7]{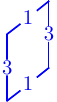}} , &&\raisebox{-.5\height}{\includegraphics[scale=0.7]{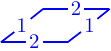}} \\
\bs A_{v,2} &= \raisebox{-.5\height}{\includegraphics[scale=0.7]{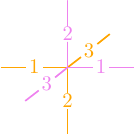}} & \bs A_{v,2}^2 &= \raisebox{-.5\height}{\includegraphics[scale=0.7]{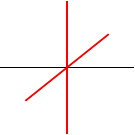}} & \bs B_{p,2}&= \raisebox{-.5\height}{\includegraphics[scale=0.7]{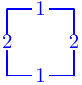}}, &&\raisebox{-.5\height}{\includegraphics[scale=0.7]{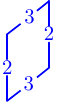}} , &&\raisebox{-.5\height}{\includegraphics[scale=0.7]{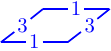}}\\
\bs A_{v,3} &= \raisebox{-.5\height}{\includegraphics[scale=0.7]{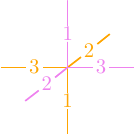}} & \bs A_{v,3}^2 &= \raisebox{-.5\height}{\includegraphics[scale=0.7]{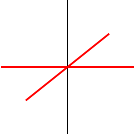}} &\bs B_{p,3}&= \raisebox{-.5\height}{\includegraphics[scale=0.7]{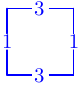}}, &&\raisebox{-.5\height}{\includegraphics[scale=0.7]{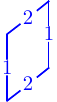}} , &&\raisebox{-.5\height}{\includegraphics[scale=0.7]{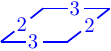}}
\end{align*}
         \begin{align*}
\bs B_{c}&= \raisebox{-.5\height}{\includegraphics[scale=0.7]{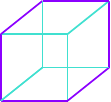}}, &\bs B_{c}^2 = \raisebox{-.5\height}{\includegraphics[scale=0.7]{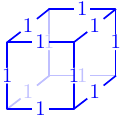}}= \raisebox{-.5\height}{\includegraphics[scale=0.7]{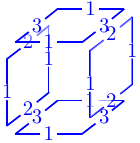}} = \raisebox{-.5\height}{\includegraphics[scale=0.7]{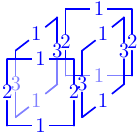}} = \raisebox{-.5\height}{\includegraphics[scale=0.7]{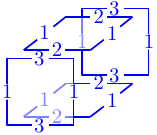}}
         \end{align*}
    \caption{{\bf Lattice Model for the Lineonic $(\ZZ_4^2,\ZZ_2^2)$ gauge theory:} Visualization of the operators in the Lineonic $(\ZZ_4^2,\ZZ_2^2)$ gauge theory. Here, ${\color{magenta}1,2,3} = IX,\cX I,\cX^\dagger X$ (magenta), ${\color{orange}1,2,3} = IX,\cX^\dagger I,\cX X$ (orange), red = $\cX^2I$, ${\color{blue}1,2,3} = \cZ^2I,IZ,\cZ^2Z$ (blue), cyan = $\cZ I$, purple = $\cZ^\dagger I$. }
    \label{fig:lineon}
\end{figure*}

In this Appendix, we introduce an isotropic version of the Lineonic Hybrid X-cube model presented in Sec. \ref{sec:Z42Z22lineon}. This model is more involved, because it is a hybrid between the X-Cube model and two 3d toric codes. Nevertheless, all the lineons in this model will square to mobile charges. Descriptively, let $e_x,e_y,e_z$ be the lineons in the X-Cube model and $e_1,\ e_2$ be the mobile charges of the toric code. Then the fusion rules are given by
\begin{align}
    e_x^2 &=e_1 & e_y^2 &= e_2, & e_z^2 &=e_1e_2 
\end{align}
Because of the gauge group underlying this model, we will call it a Lineonic $(\ZZ_4^2,\ZZ_2^2)$ gauge theory.

\subsection{Paramagnet}
First, we describe the paramagnet with a $(\ZZ_4^2,\ZZ_2^2)$ 2-subsystem symmetry. The paramagnet is on cubic lattice with two $\ZZ_4$ qudits per vertex
\begin{align}
    H=-\sum_{v}\frac{1+ \cX I_v + \cX^2 I_v +\cX^3 I_v}{4} \frac{1+ I\cX_v  +I \cX^2_v  +I \cX^3_v}{4}
\end{align}
The global $\ZZ_4^2$ symmetry has three $\ZZ_4$ subgroups of interest, generated by $\prod_v \cX I_v$, $\prod_v I\cX^\dagger_v$, and the diagonal $\prod_v \cX^\dagger \cX_v$. Furthermore, there are three $\ZZ_2$ planar symmetries $\prod_{v \in xz} \cX^2I_v$, $\prod_{v \in yz} I\cX^2_v$, $\prod_{v \in xy} \cX^2\cX^2_v$. The product of all planar symmetries in a particular direction is respectively the $\ZZ_2$ subgroup of the three $\ZZ_4$ global symmetries.

The operators that create charged excitations that commute with the above symmetry are defined on each edge $e_\rho$, where $\rho =x,y,z$ and depends on the direction of the edge.
\begin{align}
\Delta_{e_x} &= (\cZ I)_i \ (\cZ^\dagger I)_f\\
\Delta_{e_y} &=  (I \cZ)_i \ (I \cZ^\dagger)_f\\
\Delta_{e_z} &= (\cZ^\dagger\cZ^\dagger)_i \ (\cZ \cZ)_f
\end{align}
We note that the convention of the generators and hopping operators above have been carefully chosen such that the end point $i$ of $\Delta_{e_x}$ is charged $1,i,-i$ under the three $\ZZ_4$ generators, and charged $1,-1,-1$ under the planar $\ZZ_2$ symmetries. This makes the charges at the endpoint a lineon excitation. Furthermore, the product $\Delta_{e_x}\Delta_{e_y}\Delta_{e_z}$ which shares the same endpoint $i$ has no charge at site $i$. In the corresponding gauged model, this means that the three lineons mobile in the three directions fuse to the vacuum.

\subsection{Hybrid Order}
The hybrid model is defined on a 3d cubic lattice. For each edge, we place a $\ZZ_4$ qudit and a $\ZZ_2$ qubit. The Hamiltonian of the hybrid model is
\begin{align}
    H_\text{Hybrid} =& -\sum_{r=x,y,z} \left [\sum_v \frac{1+ \bs A_{v,r}+ \bs A_{v,r}^2 + \bs A_{v,r}^3}{4} \right. \nonumber\\
    &\left.+ \sum_{p} \frac{1 + \bs B_{p,r}}{2} \right]-  \sum_{c} \frac{1+\bs B_{c}+ \bs B_{c}^2+ \bs B_{c}^3}{2},
\end{align}
where
\begin{align}
\bs A_{v,r} &=\prod_{e\rightarrow v}\bs \xi_{e,r}^\dagger \prod_{e \leftarrow v}\bs \xi_{e,r},\\ 
\label{equ:Avlineon}
\bs A_{v,r}^2 &=\prod_{e \supset v_r} (\cX^2 I)_e,\\ 
\bs B_{p,r} &=\prod_{e \in p} \bs \zeta_{e,r} 
\label{equ:Bnablalineon}\\
\bs B_{c}&= \prod_{e \in c'} (\cZ I)_e \prod_{e \in c} (\cZ I)_e.
\label{equ:Bclineon}
\end{align}
Here, $e\rightarrow v$ and $e \leftarrow v$ denote edges entering and exiting the vertex $v$ as in the main text. This is shown in Figure \ref{fig:lineon} as orange and magenta respectively. Furthermore, $e \in c'$ and $e \in c$ refer to the purple and cyan edges of a cube $c$ in the Figure.

To define $\bs \xi_{e,r}$ and $\bs \zeta_{e,r}$, we note that its definition depends on the orientation of the link $e$. Therefore, we have to define it separately for $e_x$, $e_y$, and $e_z$. For simplicity in defining these operators, we substitute $x,y,z$ with $1,2,3$ so that we can define
\begin{align}
    \bs \xi_{e_{\rho},r} &= \begin{cases}
    IX &; r-\rho \equiv 0 \ (\text{mod} \ 3)\\
    \cX I &; r-\rho\equiv 1 \ (\text{mod} \ 3)\\
    \cX^\dagger X&; r-\rho \equiv 2 \ (\text{mod} \ 3)\\
    \end{cases}
    \label{equ:xilineon}\\
    \bs \zeta_{e_{\rho},r} &= \begin{cases}
    \cZ^2I &; r-\rho\equiv 0 \ (\text{mod} \ 3)\\
    IZ &; r-\rho\equiv 1 \ (\text{mod} \ 3)\\
    \cZ^2Z&; r-\rho \equiv 2 \ (\text{mod} \ 3)\\
    \end{cases}
    \label{equ:zetalineon}
\end{align}
Note that for a fixed edge $e$, $\bs \xi_{e,x} \bs \xi_{e,y} \bs \xi_{e,z}=1$, and $\bs \zeta_{e,x} \bs \zeta_{e,y} \bs \zeta_{e,z}=1$. In particular this implies that $ \bs A_{v,x}\bs A_{v,y}\bs A_{v,z}=1$. Furthermore, we would like to point out that $\bs \xi_{e,r}$ and $\bs \zeta_{e,r'}$ commutes for $r=r'$ and otherwise anticommutes.

We can see that the above model is a hybrid between two toric codes and the X-Cube model in the following way. Suppose the terms $\bs \xi_{e,r}$ and $\bs \zeta_{e,r}$ were instead Pauli matrices
\begin{align}
    \bs \xi_{e_{\rho},r} &= \begin{cases}
    IX &; r-\rho \equiv 0 \ (\text{mod} \ 3)\\
    X I &; r-\rho\equiv 1 \ (\text{mod} \ 3)\\
    X X&; r-\rho \equiv 2 \ (\text{mod} \ 3)\\
    \end{cases}\\
    \bs \zeta_{e_{\rho},r} &= \begin{cases}
    ZI &; r-\rho\equiv 0 \ (\text{mod} \ 3)\\
     IZ &; r-\rho\equiv 1 \ (\text{mod} \ 3)\\
    ZZ&; r-\rho \equiv 2 \ (\text{mod} \ 3)\\
    \end{cases}
\end{align}
Then the pair $\bs A_{v,r}$ and $\bs B_{p,r}$ forms two copies of the toric code (note that there are three terms for $r=x,y,z$ but only two are independent). Furthermore, the pair $\bs A_{v,r}^2$ and $\bs B_c$ are stabilizers for the X-Cube model. However, by promoting the first variable into a $\ZZ_4$ variable, the models are coupled in such a way that the vertex term of each toric code now squares to the vertex term (that detects the lineon) in the X-Cube model, and the cube term which detects the fracton squares to a product of plaquette terms of the two toric codes. This is illustrated in Figure \ref{fig:lineon}.

\subsubsection{Summary of excitations, fusion and braiding}
Because the model is exactly solvable, we can explicitly write down the excitations . The operator $\cZ I$ on a rigid string in the direction $\rho$ violates the vertex terms $\bs A_{v,r}$ for $r\ne \rho$. In particular, the end points are charged $\pm i$ under the operator $\bs A_{c,r}$, and $-1$ under $\bs A_{v,r}^2$, and are therefore lineons. It commutes with all $\bs B_{c_r}$ and $\bs B_{\nablapic}$'s. We will label the corresponding lineon $\bs e_{\rho}$, where $\rho =x,y,z$ is the direction of the rigid string $L_\rho$.
\begin{align}
    \text{lineon}~\bs e_\rho,\bar{ \bs e}_\rho:~~~\prod_{e_\rho\in L_\rho} {\cZ I}_{e_{\rho}}.\label{eq:elineon}
\end{align}

Similarly, the end points of $\cZ^2 I$ creates excitations which are charged $-1$ under two of the three $\bs A_{v,r}$ operators. so the point excitation $\bs e_{\rho}^2$ is created.  However, this excitation is mobile. To hop $\bs e_{\rho}^2$ in the direction $r$ we use a flexible string $L$ of $\bs \zeta_{e,r}$ in Eq. \eqref{equ:zetalineon}. Note that in for $r=\rho$, $\bs \zeta_{e,r}$ is just equal to $\cZ^2 I$ as expected.
\begin{align}
    \text{mobile}~\bs e_{\rho}^2:~~~ \prod_{e \in L}\bs \zeta_{e,r},\label{eq:e2lineon}
\end{align}

Next, we define operators that violate $\bs B_{p,r}$ and $\bs B_{c}$, but commute with $\bs A_{v,r}$. 
The first kind are loop excitations labeled $\bs m_r$ for $r=1,2,3$, and satisfy $\bs m_1 \times \bs m_2 \times  \bs m_3 =1$. The loop $\bs m_r$ is created by acting with $\bs \xi_{e,r}$ on all edges intersecting a given surface $\mathcal S'$ on the dual lattice. This creates a loop excitation at the boundary of that surface. Without loss of generality, let us choose the $\mathcal S'$ to intersect the edges in a single direction $\rho$. Because of the commutation relations between $\bs \xi_{e,r}$ and $\bs \zeta_{e,r'}$, the loop excitation are charged $-1$ under $\bs B_{p,r'}$ at the boundary of $S$ if $r\ne r'$. Furthermore, $\bs \xi_{e_\rho,r}$ commutes with $(\cZ I)_{e_{\rho}}$ if $r=\rho$, otherwise they commute up to a phase $\pm i$. The result of this is that the corners of the loop excitation are charged $\pm i$ under $\bs B_c$ only for $r \ne \rho$. In other words, for a given $r$, the loop excitation $m_r$ are charged under $\bs B_c$ in two of the three directions.
\begin{align}
    \text{mobile loop}~\bs m_r:~~~\prod_{e\perp \mathcal S} \bs \xi_{e,r}, \label{eq:mrloop}
\end{align}

Lastly, the fracton excitation is a violation of $\bs B_c$. Four fractons can be created at the corners of the operator $(\cX^2 I)_e$.
\begin{align}
    \text{fracton}~\bs m^2:~~~ \prod_{e \perp \mathcal S}(\cX^2 I)_e
\end{align}

The fusion rules can be seen from the explicit form of the operators. The lineons $\bs e_\rho$ mobile in the $\rho$ direction fuse into $\bs e_\rho^2$ which are mobile particles.

\begin{figure}
    \centering
    \includegraphics[scale=0.3]{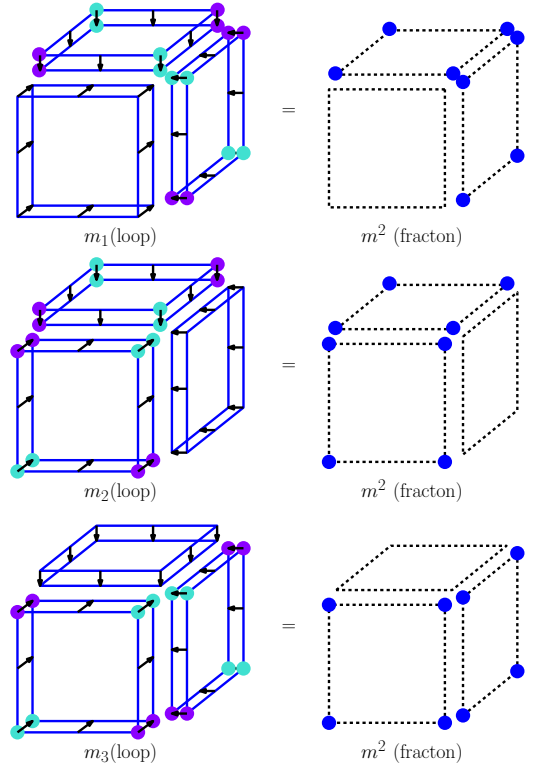}
    \caption{{\bf Loop fusion in the Lineonic $(\ZZ_4^2,\ZZ_2^2)$ gauge theory:} Fusion of two identical $\bs m_r$ loops results in fracton excitations ($\bs m^2$) at the corners if at least one of the two loop segments meeting at the corner points in the $r$ direction. }
    \label{fig:mloopfusionlineon}
\end{figure}

On the other hand, the loop excitation $\bs m_r$ is mobile. However, we notice that $\bs \xi_{e_\rho,r}^2$ is equal to $(\cX^2 I)_e$ for $r\ne \rho$. Therefore, the loop excitation fuses with itself to fractons at the corner in two of the three directions as shown in Figure \ref{fig:mloopfusionlineon}.

The braidings that differ from a stack of two toric codes and an X-Cube model is a braiding between the lineon $\bs e_\rho$ and the loop $\bs m_r$. Using the commutation relations of $\cZ I$, which creates the lineon and  $\bs \xi$ which creates the loop, one finds that the braiding phase is
\begin{align}
    \begin{cases}
    1 &; r-\rho \equiv 0 \ (\text{mod} \ 3)\\
    i &; r-\rho\equiv 1 \ (\text{mod} \ 3)\\
    -i&; r-\rho \equiv 2 \ (\text{mod} \ 3)\\
    \end{cases}
\end{align}

\section{Hybrid Haah's Code as a Parent Order}\label{app:ParentHaah}

In this Appendix, we give an identical argument to Section \ref{sec:parent} that the hybrid Haah's code is a parent state for both the toric code and Haah's code.

We add the following  perturbations to the Hybrid Haah's code Hamiltonian $H_\text{Hybrid}$ in Eq. \eqref{equ:Z4Z2Haah}.
\begin{align}
    H=H_\text{Hybrid}-t_{e^2}\sum_{e}Z_e - t_{m^2} \sum_{\alpha=1,2}\sum_v X_v^{(\alpha)},
    \label{equ:phasediagramHaah}
\end{align}

First, we derive the effective Hamiltonian for  $t_{e^2}=0$ and $t_{m^2}\rightarrow \infty$ which is the condensate of the fractonic fluxes $m^2$ by setting $X_v^{(\alpha)}=1$ for $\alpha=1,2$, .  We discard $\bs B_{v}$ since it brings us out of the subspace. The other terms in the Hamiltonian reduce to

\begin{align}
    \bs A_v &\rightarrow \prod_{e \supset v} X_e \\
    \bs B_{\nablapic} &\rightarrow \bs B_{\nablapic} = \prod_{e \in \nablapic} Z_e \nonumber \\
    \bs A_{v}^{XC} &\rightarrow 1  \nonumber
\end{align}
Therefore the effective Hamiltonian in this subspace has stabilizers of the 3d toric code.

Next, we consider $t_{m^2}=0$ and $t_{e^2}\rightarrow \infty$, which is the condensate limit of the mobile charge $e^2$. By restricting to the subspace where $Z_e=1$, the operator $\bs A_v$ is discarded, since it brings us out of the subspace. The remaining stabilizers reduce to
\begin{align}
    \bs B_{\nablapic} &\rightarrow 1\\
    \bs A_{v}^{HC} &\rightarrow X^{(1)}_{v}X^{(1)}_{v-\hat x}X^{(1)}_{v-\hat y}X^{(1)}_{v-\hat z}X^{(2)}_{v}X^{(2)}_{v-\hat x-\hat y}X^{(2)}_{v-\hat y -\hat z}X^{(2)}_{v-\hat z -\hat x} \nonumber  \\
    \bs B_{c,r} & \rightarrow Z^{(1)}_{v}Z^{(1)}_{v+\hat x+\hat y}Z^{(1)}_{v+\hat y +\hat z}Z^{(1)}_{v+\hat z +\hat x}{Z^{(2)}_{v}}{Z^{(2)} _{v+\hat x}} {Z^{(2)}_{v+\hat y}}{Z^{(2)}_{v+\hat z}} \nonumber
\end{align}
Therefore, the remaining stabilizers are those of Haah's code.
\bibliography{references}

\end{document}